%% file: main.tex
\documentclass{article}
\usepackage{graphicx} 
\usepackage[a4paper, total={6in, 8in}]{geometry}

\usepackage[bookmarksnumbered = true, colorlinks = true, linktocpage = true, bookmarksopen = true,
    citecolor=blue,%
    filecolor=blue,%
    linkcolor=blue,%
    urlcolor=blue,
	backref=page]{hyperref}

\usepackage[sort&compress]{natbib}

\usepackage{float}

\usepackage{amsfonts} 
\usepackage{amsmath}
\usepackage{bm}

\usepackage{todonotes} 
\usepackage{comment} 

\usepackage{algorithm2e}

\usepackage{epstopdf}
\AppendGraphicsExtensions{.tif}

\usepackage{xcolor}

\usepackage{svg}

\usepackage{makecell}
\usepackage{multirow} 

\usepackage{lscape}

\usepackage{lineno} 

\begin{document}

\input{titlepage}

\newpage
\section{Introduction}
Predicting the potential effects of human intervention in an ecosystem is near impossible without quantitative modelling tools \citep{baker2017_EEM,adams2020_TS,Possingham_2001,tulloch_2020}. The flow-on effects of conservation can be incredibly challenging to foresee because an effect on one species can have a chain reaction for many others \citep{mcdonald_2016}. Hence, quantitative models are essential for understanding the risk of unintended consequences \citep{baker2017_EEM,sutherland_2006} and for deciding between potential actions \citep{dexter_2012}. 

Yet, a major challenge when constructing quantitative ecosystem population models is parameterisation. The values of model parameters can have a strong influence on the population dynamics predicted by a model, so these values must be carefully selected to ensure the model predictions are reasonable. Yet, there is often little knowledge available to inform parameter values \citep{baker2017_EEM,Geary_2020}. In addition, data that can be used to constrain population predictions is often noisy, sparse, or unavailable \citep{Mouquet_2015,Novak_2011}. Critically, without data, modellers must make assumptions about how an ecosystem functions as a proxy for population data. 

Coexistence and stability are ubiquitous concepts in theoretical ecology \citep{Cuddington_2001,Allesina_2012_stab,Rohr_2014,Grilli_2017_FS,Dougoud_2018,landi_2018_review}, that are commonly used to parameterise ecosystem models \citep{baker2017_EEM,Bode_2017,Pesendorfer_2018_egEEM,Peterson_2021,Rendall_2021_EEMeg,Peterson_2021_DirkHartog,holland_2023_thesis,vollert_2023_SMCEEM}. Ecosystem \textit{stability} suggests that an ecosystem can recover from small changes to species abundances \citep{donohue_2016_stability}, and \textit{feasibility} (a concept aligned with persistence and coexistence) suggests that steady-state populations must be positive, as a negative abundance would be meaningless \citep{gravel_2016_stabilitycomplexity,song_2018_feasibility,Grilli_2017_FS}. Together, these equilibrium conditions form the ``balance of nature'' ideal, which suggests that long term ecosystem populations will stabilise towards a balance of coexisting species \citep{Cuddington_2001}. 

These concepts have been highly useful for parameterising ecosystem models in the absence of data \citep{baker2017_EEM,vollert_2023_SMCEEM}, and have been used in practice to assess conservation actions such as species reintroduction \citep{baker2017_EEM,Pesendorfer_2018_egEEM,Peterson_2021_DirkHartog}, invasive species management \citep{Rendall_2021_EEMeg,Bode_2017,holland_2023_thesis}, habitat restoration \citep{Bode_2017}, and species translocation \citep{Peterson_2021}. However, feasible and stable equilibria cannot be directly observed in an ecosystem (since they are a product of the modelling), and are therefore a challenge to verify through empirical evidence \citep{francis_2021,reimer_2021}. Here, along with a growing body of literature, we question whether this idealistic representation of ecosystems is appropriate, particularly in the context of conservation \citep{Hastings_2018_transient,Cuddington_2001,francis_2021,boettiger_2021,wallington_2005,Mori_2011,morozov_2020,oro_2023}. 

Feasibility and stability describe ecosystem behaviour near equilibrium, such that population dynamics are not affected by strong abiotic or biotic factors \citep{botkin_1990}. In the modern world, where climate change, human development, and invasive species are persistent threats across the globe, ecosystems are rarely unaffected by strong external factors \citep{waltner_2003,wallington_2005}. If ecosystems in nature are disturbed by strong external influences, then it is inappropriate to model them at an equilibrium state  \citep{deangelis_1987, Mori_2011, wallington_2005}. 

While ecosystems may appear to be exhibiting steady-state equilibrium behaviour, this could be a mismatch between human and ecological timescales \citep{francis_2021}. Ecosystems may exhibit long transient dynamics, such that non-equilibrium, slow-changing populations are observed \citep{Hastings_2018_transient,francis_2021,morozov_2020,boettiger_2021}. An ecosystem represented at equilibrium when compared to one represented as a long transient can yield qualitatively different model parameterisations \citep{Hastings_2001_transient}, and different model-informed conservation decision making \citep{oro_2023,boettiger_2021}. Yet, it is a challenge to even identify if an ecosystem is experiencing long transient or equilibrium behaviour \citep{boettiger_2021,reimer_2021}.

Despite these criticisms, equilibrium perspectives remain pervasive in ecological modelling \citep{morozov_2020,oro_2023}. To our best knowledge, there is currently no method for constraining ecosystem model behaviour that does not rely on time series data or equilibrium assumptions. Since population monitoring data is often sparse, noisy or unavailable \citep{Mouquet_2015,Kristensen_2019,Geary_2020,baker2017_EEM}, the equilibrium perspective is therefore needed to represent ecosystems without investing significant time and money into data collection. 

Alternatively, we propose a framework for constraining model outputs by removing any parameter values that lead to obviously unrealistic behaviour, according to expert knowledge and field observations. To illustrate this concept, we demonstrate a series of four approaches to constraining model population predictions, starting with the classic feasible and stable requirements and increasing in realism. Firstly, we introduce the option to restrict population sizes to within expert-elicited limits, such that models that yield unrealistic equilibrium populations are excluded. Secondly, instead of analysing equilibrium populations, we analyse population trajectories; this equilibrium-agnostic alternative addresses the likely common situation of transient (non-equilibrium) ecosystem dynamics \citep{Hastings_2001_transient}. Finally, we introduce the option to limit population fluctuations based on field knowledge, to restrict populations from growing or shrinking at unobserved or extreme rates. This new framework allows ecosystems to be represented without relying on an equilibrium paradigm or data availability, by instead utilising any available knowledge of historical populations. 

To implement this new framework, we develop a novel approach to generating ensembles of model parameters whose predictions match the expected behaviours. Approximate Bayesian algorithms \citep{sunnaaker_2013_ABC} provide a path for parameterising using non-data constraints, but new algorithms are needed to parameterise these models in an efficient manner. Here, we significantly extend an approximate Bayesian sequential Monte Carlo algorithm \citep{vollert_2023_SMCEEM} to rapidly parameterise time-series trajectories where the previous approach is too computationally infeasible to consider.

Our new framework, together with statistical advances that make it practically accessible, use far more pragmatic assumptions to construct ecosystem models based on the available knowledge of a system. To our best knowledge, this is the first equilibrium-agnostic framework for representing ecosystems without data. To do so, we maximally utilise expert knowledge and/or field observations -- a readily available and often overlooked source of information. We show that this new framework has the potential for dramatic impact on the management of complex ecosystems, improving both the precision of the models that represent these ecosystems and the confidence in any forecasted scenarios in these ecosystems.

\section{Methods}

\subsection{Assumptions for constraining ecosystem models}
\label{Methods: constraint sets}

In the present work, an ecosystem model refers to any model that can simultaneously forecast multiple populations through time. There are a multitude of suitable modelling frameworks, from deterministic ordinary differential equations (e.g.,\ generalised Lotka-Volterra; \cite{adams2020_TS,baker_2019}), to individual-based, spatial, or stochastic models (see \cite{Geary_2020} for a recent review of ecosystem models). Where the chosen model parameters affect the resulting population estimates, we constrain their parameter values such that they must lead to reasonable population dynamics. Here, we outline four sets of assumptions that could be used for constraining ecosystem population models and summarise these in Table \ref{tab:constraint summary} (Supplementary Material Section S.1 contains mathematical descriptions of each set of constraints).  

\begin{table}[!ht]
    \centering \small
    \begin{tabular}{p{0.25\linewidth} |p{0.7\linewidth}}
         \textbf{Set of constraints} & \textbf{Description} \\ \hline \hline
          (1) Positive and stable equilibria  
             & In the long term, populations will coexist and be able to recover from small changes. These constraints are commonly considered in the literature (e.g.,\ \citet{baker2017_EEM,Allesina_2012_stab,song_2018_feasibility}).  
             \\ \hline
          (2) Bounded and stable equilibria 
             & In the long term, populations will be bounded and be able to recover from small changes. Here, equilibrium abundances are restricted to expert elicited ``reasonable'' limits. 
             \\ \hline
          (3) Bounded trajectories
             & For a specified period of time, the population sizes will be bounded. Here, we only consider short-term non-equilibrium behaviour. 
             \\ \hline
          (4) Bounded and slow-changing trajectories 
             & For a specified period of time, the population sizes will be bounded and are limited by how rapidly they can increase or decrease. Here, any expert-elicited ``unreasonable'' population changes are excluded. 
    \end{tabular}
    \caption{A summary of the sets of constraints used to generate parameter sets for ecosystem modelling. }
    \label{tab:constraint summary}
\end{table}

\subsubsection{Equilibria are feasible and stable}
Firstly, we consider the classic feasible and stable equilibrium constraints \citep{baker2017_EEM,Bode_2017,Pesendorfer_2018_egEEM,Peterson_2021,Rendall_2021_EEMeg,Peterson_2021_DirkHartog,Allesina_2012_stab,Dougoud_2018,donohue_2016_stability,Grilli_2017_FS}. To assess whether a model meets these constraints, the equilibrium populations must be positive and locally and asymptotically stable. 

The resulting model predictions of a feasible and stable model will tend toward positive populations, such that, population trajectories will eventually become attracted to the positive fixed point, or they will oscillate around these equilibrium values \citep{deangelis_1987,Edelstein-Keshet_2005_mathbio}. Hence, these models represent idealised long-term behaviour of ecosystems in the absence of any large external impacts: an ideal known as the ``balance of nature'' \citep{Cuddington_2001}. However, these constraints place no restriction on the size of equilibrium populations and can yield extremely low or high equilibrium abundances. 

\subsubsection{Equilibria are bounded and stable}
Feasibility requires equilibrium populations to be positive, but we may wish to further restrict the possible long-term abundances; for example, to enforce physical limits on vegetation biomass or to require a minimum invertebrate biomass for sustaining an ecosystem. In addition to feasible and stable equilibria, this second set of constraints requires that equilibrium populations are within specified bounds for a subset of the species, where any unspecified bounds limit equilibria to positive values (feasibility).  

Much like models constrained by feasible and stable equilibria, the resulting population trajectories for this set of constraints will stably approach or oscillate around a fixed point in population space in the long term. However, in this case, the fixed point is not only forced to be positive (feasible), but also within ``reasonable'' population limits, as informed through expert elicitation. 

This set of constraints may be advantageous to simply assuming feasibility and stability because it further grounds population sizes to reality. However, as this is an equilibrium-based constraint, only the little-known long-term behaviour of the system is restricted by this constraint, hence unreasonable populations may still be modelled as the system approaches equilibrium.  

\subsubsection{Trajectories are bounded}
To avoid enforcing the equilibrium perspective, we next consider finite population trajectories rather than the equilibrium behaviour of the system. The third constraint set requires population trajectories that remain within specified bounds for a finite period of time $T$, which we refer to as the observation period.

This constraint draws on field observations to determine the minimum and maximum possible populations for a period of observation (e.g., \textit{``in the last 10 years there were 200-600 dingoes in the national park''}). This approach assumes finite coexistence using historical beliefs (irrespective of whether the ecosystem is at equilibrium), rather than assuming species will indefinitely persist in an equilibrium paradigm. Hence, feasibility is enforced for the observation period $T$, but beyond this extinctions are possible.

There are many arguments for why non-equilibrium approaches may be advantageous (see e.g.,\ \citet{oro_2023,Hastings_2018_transient,francis_2021}), and this set of constraints allows modellers to parameterise non-equilibrium behaviour even without data. Since the stability of the system is no longer be assessed (as the equilibrium behaviour is not considered), this set of constraints cannot limit populations from rapidly crashing or exploding.

\subsubsection{Trajectories are bounded and slow-changing}
The last set of constraints we demonstrate imposes restrictions on the rate of population change within the observation period to limit how rapidly populations can rise or fall. Here, we restrict population trajectories for the period $T$, such that both population sizes and the population rate of change are within specified bounds. 

Again, this approach aims to utilise field observations (e.g., \textit{``for the last 5 years, populations never halved or doubled in a month''} and expert knowledge of the ecosystem (e.g.,\ limiting population growth based on reproductive capabilities). Beyond the constrained period, species may go extinct or have explosive populations, hence we can think of this constraint as finite coexistence with ``slow-changing'' or ``pseudo-stable'' behaviour, which may be in line with long transients or stable equilibrium dynamics. 

\subsection{Generating samples that meet the constraints}
\label{Methods:sample generation}
For any given model structure, we aim to obtain a \textit{representative} ensemble of parameter sets that meet the constraints (such that the ensemble provides a reasonable and well-balanced sample of all areas of parameter space that meet the constraints). Since this parameterisation process must be done without data, standard parameter inference techniques that require data cannot be applied. Instead, we use  ideas from approximate Bayesian inference, where datasets are turned into summary statistics that are used for inference \citep{sunnaaker_2013_ABC}; instead, in our case, the set of ecosystem constraints that must be satisfied are treated as summary statistics \citep{vollert_2023_SMCEEM}. To our best knowledge, only two approximate Bayesian sampling algorithms have been used for generating ensembles of ecosystem models to meet theoretical assumptions of ecosystem models. Firstly, an accept-reject algorithm that rejects any sampled parameter sets that do not satisfy all constraints has been used to find feasible and stable ecosystem models \citep{baker2017_EEM}, but this approach can be slow when there is a low probability of randomly sampling appropriate parameter values \citep{vollert_2023_SMCEEM}. Instead, the Sequential Monte Carlo ensemble ecosystem modelling (SMC-EEM) algorithm can be far more efficient at parameterising high-dimensional ecosystem networks to satisfy feasibility and stability constraints, while still producing a representative ensemble \citep{vollert_2023_SMCEEM}. Thus, below we summarise the SMC-EEM method, then introduce a new temporally adaptive extension of this method that provides an elegant solution for trajectory-based constraints. 

\subsubsection{SMC-EEM}
The SMC-EEM algorithm uses information from rejected parameter sets to sequentially propose new values from a more informed distribution of potential parameter sets, rather than always using the prior distribution \citep{vollert_2023_SMCEEM}. The general idea is to iteratively reduce the measured ``discrepancy'' between the model simulations and ecosystem assumptions until the resulting ensemble of models all fully satisfy the enforced ecosystem constraints. 
For further details of the algorithm see \citet{vollert_2023_SMCEEM}, and implementations are available in MATLAB \citep{Vollert_2024_code} and R \citep{Pascal_2024_code}. SMC-EEM is an efficient choice when the discrepancy is evaluated by analytically calculating the equilibrium (as in constraint sets (1) and (2) in Table \ref{tab:constraint summary}); however, simulating full time series trajectories can add significant computation time, making SMC-EEM too slow to consider when trajectory-based constraints (sets (3) and (4) in Table \ref{tab:constraint summary}) are present. 

\subsubsection{Temporally adaptive SMC-EEM} 
One downside of using SMC-EEM for trajectory-based constraints is that even if a trajectory has undesirable behaviour (and a very high discrepancy) in the early stages of the simulation, the full period $T$ is simulated, leading to wasted computation time. Instead, we propose a temporally adaptive modification of the SMC-EEM algorithm that avoids simulating the full time series by sequentially increasing, at each iteration $k$, the simulation time $t_k \leq T$, as more and more parameter sets are found that meet the constraints for the simulated period. Our new temporally adaptive SMC-EEM algorithm simultaneously decreases the discrepancy of model simulations from the constraints in the period $[0, t_k]$ and increases the simulation period $t_k$ towards the full time period $T$. Hence, we introduce and use the temporally adaptive SMC-EEM algorithm to generate ensembles of parameter sets for trajectory-based constraints in the present work (sets (3) and (4) in Table \ref{tab:constraint summary}). For details of the temporally adaptive SMC-EEM algorithm, see Supplementary Material Section S.2.

\section{Results}
Starting from the classic equilibrium constraints -- feasible and stable equilibria (set (1) in Table \ref{tab:constraint summary}) -- we demonstrated a series of increasingly pragmatic alternatives for representing ecosystems without data (see Table \ref{tab:constraint summary} for a summary), either by constraining the long-term equilibrium behaviour (set (1) and (2)) or when considering finite population trajectories (sets (3) and (4)). We demonstrate these new pragmatic constraints for generalised Lotka Volterra models -- a common choice in the literature \citep{Dougoud_2018,baker2017_EEM,adams2020_TS} -- combined with our efficient statistical framework for generating parameter sets that satisfy the constraints. Here, our results illustrate that elicited knowledge of expected populations (sets (2)-(4)), or observed population fluctuations (set (4)) can be used in place of equilibrium assumptions to yield knowledge-derived representations of ecosystems.

\subsection{Constraint choice drastically alters species population predictions, illustrated by a three-species predator-prey system}

Each of the four sets of constraints we analysed led to qualitatively different population trajectories (as shown for a simple three-species predator-prey ecosystem in Figure \ref{fig:3_species_trajectories}; full details of the ecosystem model provided in Supplementary Materials Section S.3.1). First, if the classic feasible and stable equilibrium-based constraints (constraint set (1) in Table \ref{tab:constraint summary}) are used to generate ecosystem trajectories, the resulting steady-state populations are not restricted, and can result in extremely large or small populations that do not necessarily adhere to ``reasonable limits'' that we may wish to enforce (Figure \ref{fig:3_species_trajectories}b, row 1). Introducing limits to \textit{equilibrium} populations (constraint set (2) in Table \ref{tab:constraint summary}) can be used to force the long-term behaviour within some limits; however, populations can still exceed these before reaching equilibrium (Figure \ref{fig:3_species_trajectories}b, row 2). Instead, the non-equilibrium assumption that constrains \textit{trajectories} to be within reasonable bounds (constraint set (3) in Table \ref{tab:constraint summary}) removed any possibility of observing unrealistic population sizes for a period of interest (Figure \ref{fig:3_species_trajectories}b, row 3). Finally, in addition to constraining trajectories by population sizes, limiting the rate of population change (constraint set (4) in Table \ref{tab:constraint summary}; Figure \ref{fig:3_species_trajectories}b, row 4) prevents rapid changes in populations  that may be present in other sets of constraints where the rate of population change is not limited. Hence, the choice of constraints can drastically alter the population predictions obtained for an ecosystem model. Beyond these individual model trajectories, the constraint choice also affects predictions from an ensemble of trajectories (Figure S1) and the parameter estimates (Figure S2). 

\begin{figure}[!ht]
    \centering    
    \includegraphics[width=0.75\textwidth]{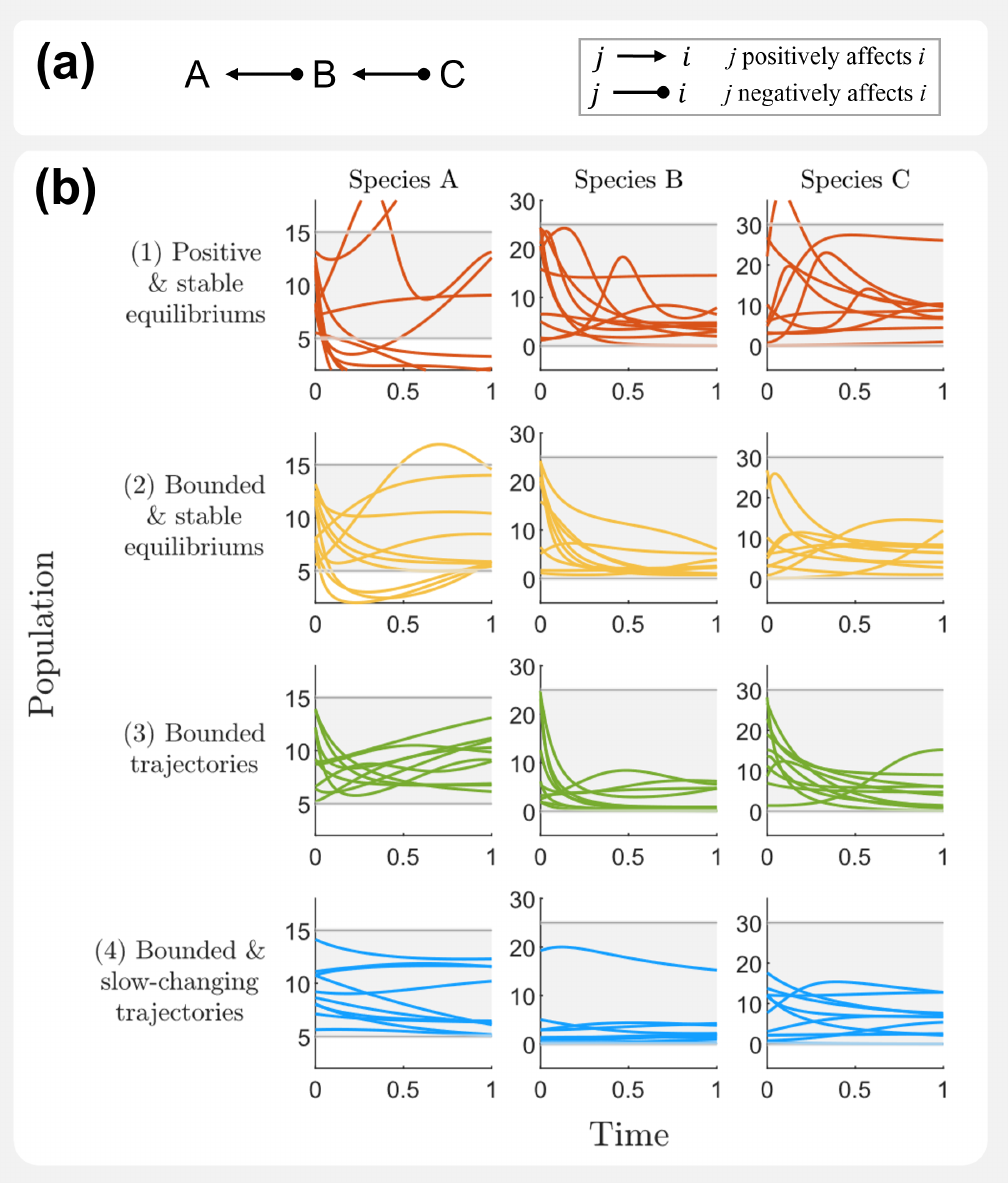}
    \caption{\textbf{(a)} A three-species predator-prey ecosystem network constructed into a Lotka-Volterra ecosystem model (see Supplementary Material Section S.3.1 for further detail). \textbf{(b)}
    Ten examples of population predictions for the three-species ecosystem model obtained using each set of constraints. The shaded area between the grey lines indicates the expert-elicited ``reasonable'' population limits (see Supplementary Material Section S.3.1 for further detail). Notice that the choice out of the four sets of constraints can lead to considerably different population predictions.}
    \label{fig:3_species_trajectories}
\end{figure}

\subsection{Constraint choice drastically alters predicted management outcomes, illustrated by an eight-node semi-arid Australian ecosystem model}

\begin{figure}[p]
    \centering 
    \includegraphics[width=\textwidth]{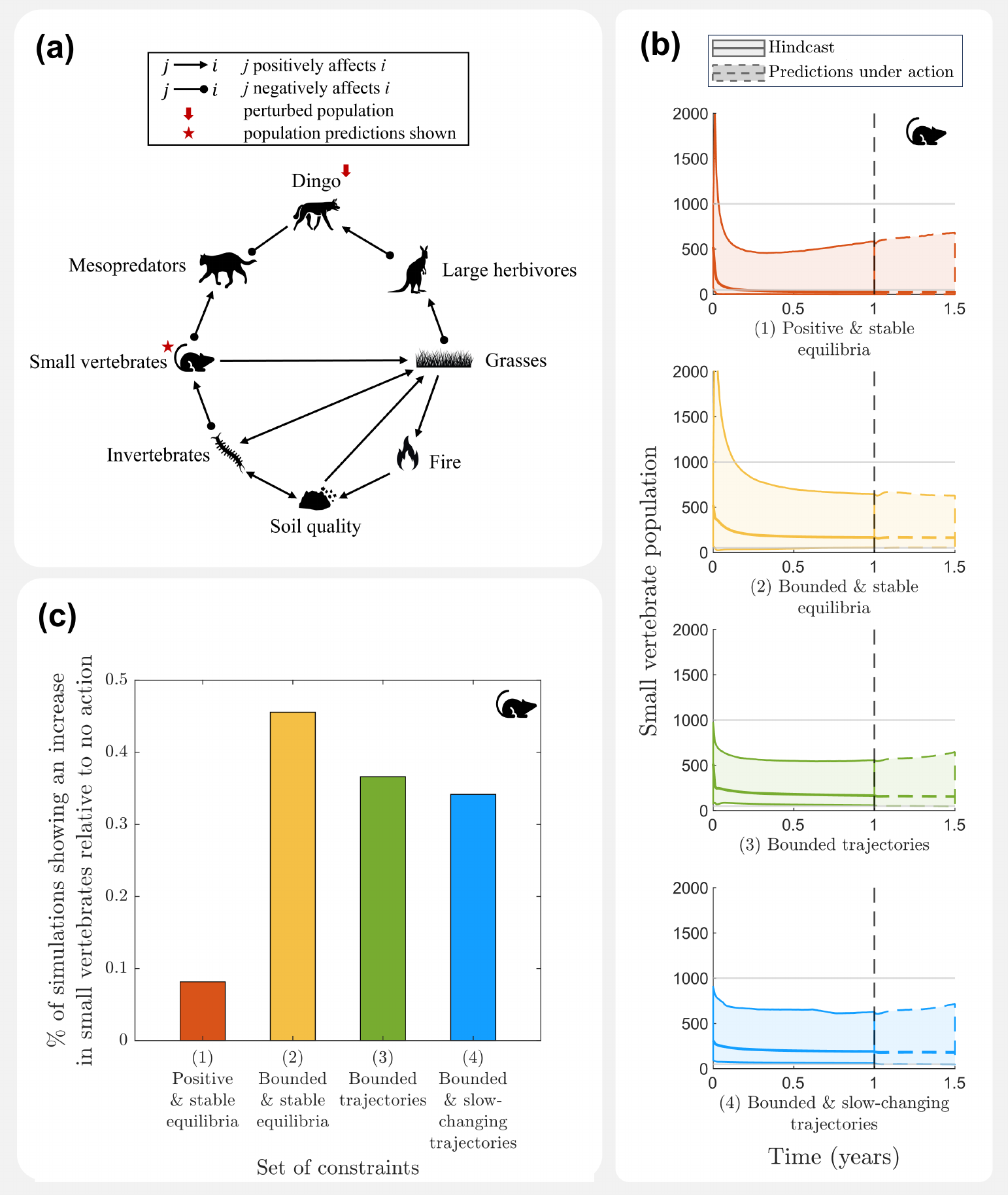}
    \caption{(Caption next page.)}
    \label{fig:dingo_decision}
\end{figure}

Vastly different population responses are obtained depending on the constraints used to model the ecosystem-wide effects of dingo regulation in semi-arid Australia (see ecosystem network in Figure \ref{fig:dingo_decision}a; \citet{newsome_2015_dingo,baker2017_EEM}). Modelled populations (e.g.,\ for small vertebrates, Figure \ref{fig:dingo_decision}b) can change considerably based on the constraint choice, both for hindcasts (left of vertical dashed line) and in response to a pulse perturbation removing 50\% of the dingoes (right of vertical dashed line). Figure \ref{fig:dingo_decision}b reveals that the modelled populations can be vastly different between constraints (e.g.,\ the 95\% prediction interval yields a maximum population of approximately 3000 for set (1), compared to 900 for set (4)), and can show qualitatively different trends in response to dingo removal (e.g.\, the 95\% prediction interval indicates that in response to the action, small vertebrate populations may rise under set (1), but fall under set (2)). Additionally, the certainty in how populations respond can change drastically based on the set of constraints chosen. For example, when predicting the difference in small vertebrate populations with and without management, an ensemble constructed using set (1) predicts decreasing populations with confidence (92\% of simulations predict decreases), whereas, the other sets of constraints are much more uncertain in the response of small vertebrates (55\%, 64\% and 66\% of simulations predict decreases for sets (2), (3) and (4), respectively; see Figure \ref{fig:dingo_decision}c). Such differences in the predicted outcomes of management could ultimately lead to different conservation decision-making. See Supplementary Material Section S.4.1 for further information on this case study.

\addtocounter{figure}{-1}
\begin{figure} [t!]
  \caption{(Previous page.) \textbf{(a)} A semiarid Australian ecosystem network simulated by a Lotka-Volterra ecosystem model (see Supplementary Material Section S.4.1 for further detail). To demonstrate a conservation decision, we simulate a pulse perturbation removing 50\% of the dingo population, and analyse the consequent effects on population predictions. For simplicity, we only show the predicted small vertebrate population responses. \textbf{(b)} Predicted small vertebrate population responses (median and 95\% prediction interval) to dingo regulation modelled using each set of constraints. Each ensemble of models, constrained by different ecosystem dynamics, was used to forecast populations for a year, then predict population responses for six months following the perturbation. Notice that the choice of constraints affects both the hindcast populations, and the response to the management action, such that the populations are orders of magnitude different and indicate different trends. \textbf{(c)} Predicted change in small vertebrate populations 35 days after dingo regulation, compared to populations predicted with no intervention (counterfactual). This figure shows the percentage of model simulations generated using each set of constraints that led to an increase in small vertebrate populations in response to the action. Notice that the certainty of the population responses varies significantly depending on the set of assumptions chosen, such that set (1) shows a very low probability that populations increase, whereas other constraint sets show high uncertainty in the small vertebrate response after 35 days.}
\end{figure}

\subsection{Relatively stable ecosystem trajectories can easily be obtained without enforcing feasible and stable equilibrium populations}

\begin{figure}[!ht]
    \centering
    \includegraphics[width=0.75\textwidth]{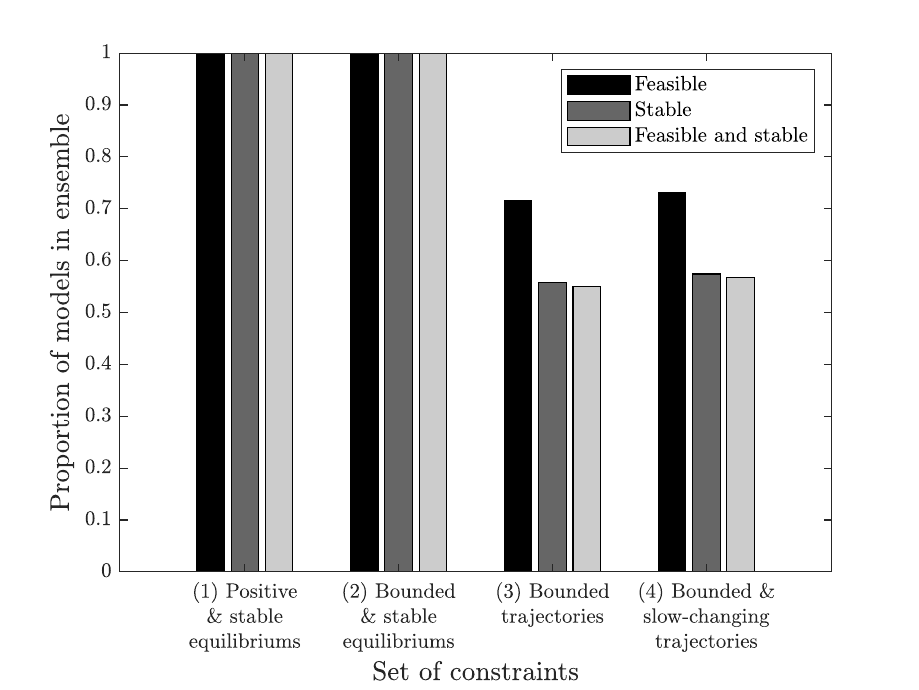}
    \caption{The proportion of parameter sets that have equilibrium behaviour that is feasible, stable or both for the semiarid Australia ecosystem model (Figure \ref{fig:dingo_decision}a; see Supplementary Material Section S.4.1 for further detail). Notice that the trajectory-based constraints lead to many parameterisations that are not feasible and stable, yet exhibit reasonable population dynamics (Figure S3).}
    \label{fig:implied feas and stab}
\end{figure}

Ensembles constrained by their non-equilibrium behaviour (constraint sets (3) and (4) in Table \ref{tab:constraint summary}) exhibit reasonable population dynamics during the period of interest (Figure S3). However, for the semiarid Australia ecosystem model (Figure \ref{fig:dingo_decision}a), roughly half of the trajectory-based ensembles (constraint sets (3) and (4) in Table \ref{tab:constraint summary}) were either infeasible or unstable (Figure \ref{fig:implied feas and stab}). Hence, we have shown that reasonable ecosystem dynamics can also be infeasible and unstable; yet, equilibrium constraints (e.g.\, sets (1) or (2) in Table \ref{tab:constraint summary}) will exclude these models. Additionally, this result indicates fundamental differences in the long-term behaviour of the models depending on the selected constraints.

\subsection{Faster ecosystem generation using a new temporally adaptive algorithm}

\begin{figure}[p]
    \centering 
    \includegraphics[width=\textwidth]{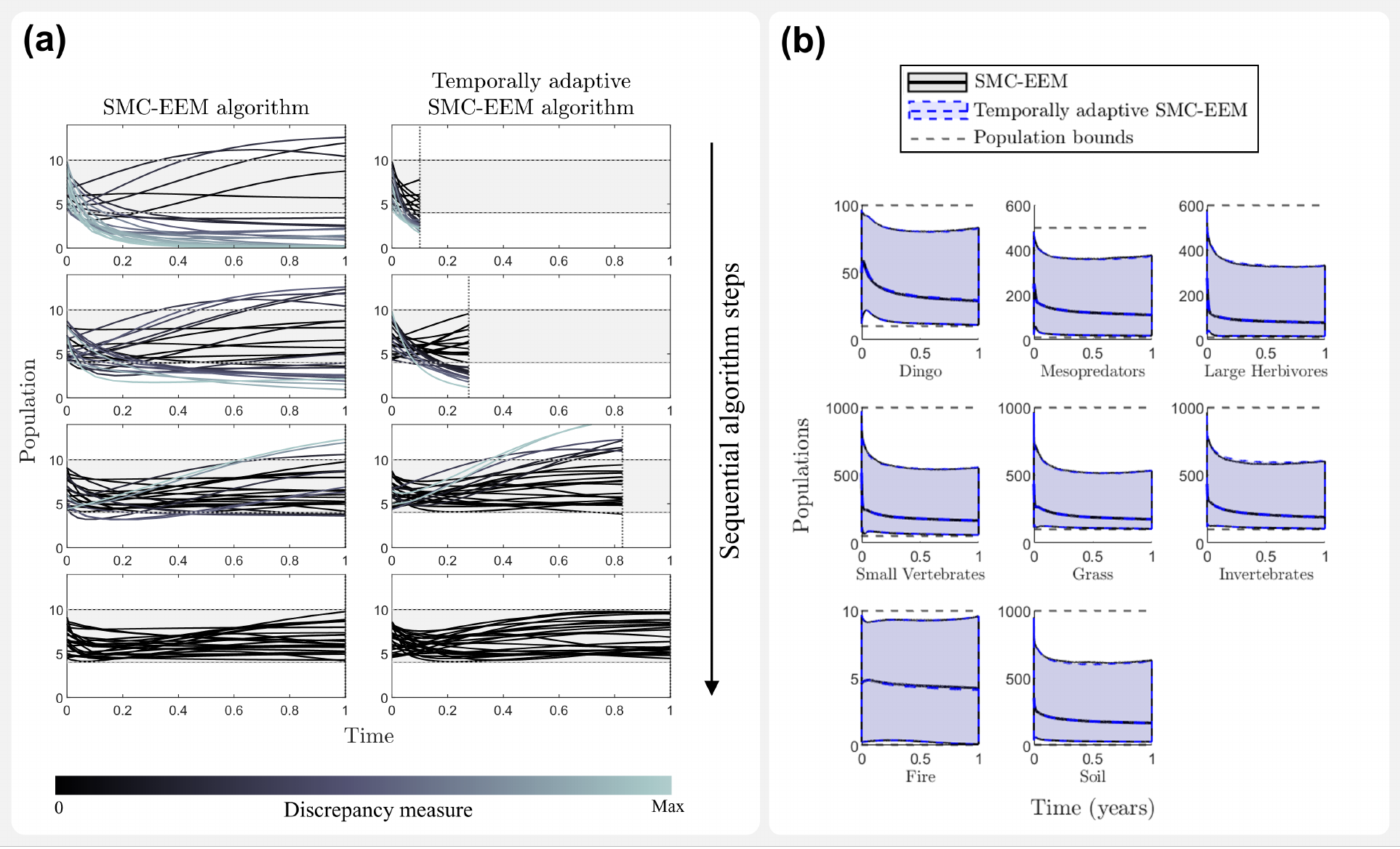}
    \caption{\textbf{(a)} A conceptual schema highlighting the differences between SMC-EEM (left) and temporally adaptive SMC-EEM (right). In this example, both algorithms aim to generate trajectories with populations in the grey area. Each simulated population trajectory has an associated discrepancy measure, ranging from no discrepancy (black; trajectories meet constraints) to the maximum observed discrepancy (light blue; trajectories furthest from meeting constraints). From top to bottom, this figure shows that both algorithms iterate the trajectories closer to meeting the constraints, by sequentially reducing the discrepancy measures. The key difference is that SMC-EEM always simulates the full period (one unit of time). On the other hand, temporally adaptive SMC-EEM initially only simulates a small period (0.1 units of time) to rapidly eliminate trajectories that do not meet the constraints, and subsequently increases the period, focusing on increasingly promising trajectories. (Caption continued on next page.)  }
    \label{fig:algorithm comparison}
\end{figure}

Generation of ecosystem parameterisations that satisfy the trajectory-based constraints can be slow even when using the most recent SMC-EEM algorithm proposed for its computational efficiency \citep{vollert_2023_SMCEEM}, in one case taking over 81 hours on a high performance computer (Table S4; 12 parallel cores). However, the computation time can be cut using a new temporally adaptive version of this algorithm introduced in the present work (Methods Section \ref{Methods:sample generation}). Put simply, our new algorithm initially simulates a shorter period when finding parameterisations that meet the constraints, and sequentially increases the simulation period until the full time period is covered (Figure \ref{fig:algorithm comparison}a). The new temporally adaptive algorithm produces equivalent ensembles to the unaltered SMC-EEM algorithm (Figure \ref{fig:algorithm comparison}b), whilst being up to four times faster (Table S4).

\addtocounter{figure}{-1}
\begin{figure} [t!]
  \caption{(Previous page.) \textbf{(b)} The population predictions for the semiarid Australian ecosystem model constrained using bounded population trajectories (set (3) in Table \ref{tab:constraint summary}; see Supplementary Materials Section S.4.1 for more information on this case study). Here we compare the median and 95\% prediction intervals generated using both SMC-EEM (black) and temporally adaptive SMC-EEM (blue, dashed). Notice that both parameterisation algorithms produce similar forecasting distributions (see Supplementary Material Section S.4.3 for further comparison), yet the temporally adaptive algorithm is up to four times faster (Table S4).}
\end{figure}

\section{Discussion}

\subsection{The argument for relaxing equilibrium assumptions when modelling ecological phenomena} 
Coexistence and stability conditions are ubiquitous in conservation modelling \citep{Rendall_2021_EEMeg,Peterson_2021_DirkHartog} and in community ecology analyses \citep{Allesina_2012_stab,Dougoud_2018,Grilli_2017_FS}, despite concerns of whether the equilibrium perspective is appropriate for ecosystems \citep{Mori_2011,oro_2023,francis_2021}. Given increasingly strong evidence and arguments against the concepts of ecological equilibrium and the ``balance of nature'' ideal \citep{Cuddington_2001}, here, we argue that coexistence and stability are inappropriate for conservation modelling when ecosystem knowledge can be used instead. In this work, we introduced and demonstrated quantitative methods for how this can be achieved. The resulting changes in projections are not trivial -- rather, drastic consequences of these assumptions are uncovered (see also Section \ref{Res: in practice}) -- so we lay out here our justifications for the re-evaluation of feasibility and stability conditions.

There is potentially a very large difference between the long-term equilibrium behaviour of an ecosystem and its current state \citep{francis_2021,morozov_2020}, and this difference can lead to significantly different representations \citep{Hastings_2001_transient} (Figures \ref{fig:3_species_trajectories}, \ref{fig:dingo_decision}, and S1-S4) and conservation practices \citep{oro_2023,francis_2021} (Figure \ref{fig:dingo_decision}b). This is of particular concern for ecosystems in need of conservation or management, due to their disturbed and potentially unstable state. These ecosystems may never reach their idealised long-term state; in fact, most time series data do not show evidence of an equilibrium state \citep{oro_2023}, or are challenging to distinguish from long transient behaviour \citep{boettiger_2021,reimer_2021,francis_2021}. Therefore, representations of these ecosystems should seek to replicate the observable system behaviours without forcing equilibrium assumptions, since many ecosystems that appear to be at equilibrium may instead be demonstrating long-transient behaviour \citep{Hastings_2018_transient,morozov_2020,francis_2021}. In this work, we echo the sentiments of a growing body of literature which suggests that it is inappropriate to only consider ecosystems at equilibrium \citep{oro_2023,Hastings_2018_transient,morozov_2020,wallington_2005,francis_2021,Mori_2011}.

\subsection{Field observations and/or expert knowledge should always fill the assumption gap}
The arguments against ecological equilibria clearly raise concern about their persistent use in the literature. However, if making conclusions generally (e.g.,\ \citet{gellner_2023}) or if data is limited (e.g.,\ \citet{Rendall_2021_EEMeg}), some assumptions about ecosystem dynamics are necessary. We assert that population data should not be estimated using equilibrium theory, but with population observations from experts. The quantitative methods presented here (summarised in Table \ref{tab:constraint summary}) provide a path for accomplishing this by using information that is expected to be available for most ecosystems. 

The new methods we introduce here utilise (when available) expert-elicited upper and/or lower limits of population abundances, and upper limits on population rates of change. Information on \textit{equilibrium} population limits (for constraint set (2) in Table \ref{tab:constraint summary}) can be elicited based on any knowledge of a species' carrying capacity or by estimating minimum sustainable populations. Similarly, population limits for an \textit{observation period} (for constraint sets (3) and (4) in Table \ref{tab:constraint summary}) can be estimated by experts (see e.g.\ \citet{Bode_2017}), even though  populations may have been outside this range beyond the observation period. Limits on population rates of change (for constraint set (4) in Table \ref{tab:constraint summary}) can be elicited from experts, where this information is informed by ecological insights such as the maximum population growth (e.g.,\ maximum eggs laid per year), or by whether sudden population changes have been observed (e.g.,\ ``\textit{for the last ten years of observation, we believe that populations have never halved over the period of a month}'').

Since these elicited quantities are being treated as data, if they cannot be confidently ascertained, conservative estimates can be provided (e.g.,\ by multiplying best estimates by a factor of ten), limits can be left unspecified for certain species, or sensitivity analyses can be performed to identify if/when an estimate affects the represented system (this may be used to decide if further data collection is necessary; see \citet{plein_2022} Fig 3 for a similar analysis). Elicitation provides a fast and cheap way to input system knowledge \citep{Bode_2017}. While this information can be imperfect -- it can contain biases \citep{martin_2012} and different experts may provide contradictory viewpoints (see e.g.,\ \citet{Peterson_2021_DirkHartog}) -- any observed information is of immense value when the alternative is to have no data \citep{Bode_2017,kuhnert_2010}. 

We acknowledge that choosing between imperfect ecological theory and imperfect elicited information may depend on a modeller's philosophy. However, it is the opinion of the authors that an expert's conservative estimate of what is impossible should be trusted above assumptions of population dynamics theorised under idealised settings.

\subsection{Towards using equilibrium-agnostic and expert-derived knowledge in practice}
\label{Res: in practice}
While the scientific consensus is gradually moving away from the equilibrium perspective \citep{moore_2009}, these ideas have not followed practice \citep{Mori_2011,wallington_2005}, due to the challenges of implementing transient theories in management \citep{francis_2021}. To our best knowledge, this is the first framework for parameterising transient ecosystem dynamics without time-series data. To illustrate our new framework, we developed three new ecosystem constraints (sets (2)-(4) in Table \ref{tab:constraint summary}) as pragmatic alternatives to feasibility and stability for ecosystem modelling (set (1) in Table \ref{tab:constraint summary}), two of which consider non-equilibrium behaviours. 

When compared the classic feasible and stable equilibrium criteria (set (1)), introducing expert-informed limits on equilibrium populations (set (2)) can make equilibrium constraints more realistic by preventing models from stabilising on unreasonably large or small populations (Figure \ref{fig:3_species_trajectories}). While this constraint restricts equilibrium populations to a reasonable domain, population dynamics are unconstrained outside of the equilibrium dynamics, that may never be reached in nature \citep{francis_2021,morozov_2020}. Therefore, for further constraints (set (3) and (4)), we instead consider population trajectories, regardless of whether they are at equilibrium or not; this choice to be equilibrium-agnostic is particularly appropriate as distinguishing equilibrium from non-equilibrium behaviour in time series data is a challenge in of itself \citep{boettiger_2021,reimer_2021}. This change to an equilibrium-agnostic perspective circumvents many of the issues associated with the equilibrium perspective: it replicates the observable ecosystem dynamics over the period the system has been managed; it allows for the possibility that the ecosystem may not be at or near equilibrium; and it does not restrict ecosystem behaviour outside the observation period (in case the observed system behaviour is e.g.,\ a long transient \citep{Hastings_2018_transient}).

Through limiting population trajectories to a reasonable population domain (set (3)), the strictness of assuming equilibrium behaviour is relaxed, such that ecosystem dynamics are reasonably constrained (Figure \ref{fig:3_species_trajectories}) without enforcing the equilibrium ideal (Figure \ref{fig:implied feas and stab}). The population limits for trajectory-based constraints (sets (3) and (4)) define the populations that could reasonably have been expected during the observation period, hence they may be specified less conservatively than when constraining the theoretical equilibrium populations (set (2)). Finally, preventing impossible or unobserved population fluctuations (set (4)) can further ground these models in reality, such that the chaotic dynamics that ecosystem models can produce are only modelled during the observation period if experts agree it was possible (Figure \ref{fig:3_species_trajectories}). Overall, this process of redefining constraints has relaxed the equilibrium assumption such that constraints are equilibrium-agnostic, whilst increasing the restrictions on reasonable population dynamics using field knowledge.

The new constraints demonstrated in this work represent a starting point to a new way of thinking, rather than a definitive list of options for constraining ecosystem dynamics. For any specific conservation scenario, knowledge beyond reasonable population sizes and changes may be available. For example, there may be knowledge of how populations have responded to management (e.g., fox control leading to greater glider extinction; \cite{baker_2019}), observed population trends (e.g., most Great Barrier Reef coral colonies have experienced major decline; \cite{death_2012}), or other ecological insights (e.g., agricultural rewilding can take 50 years to result in woodland; \cite{broughton_2021_woodland}). The framework we have developed allows systematic removal of population predictions which are known to be impossible. If the difference between expected and simulated population responses can be defined, it can be parameterised. System specific knowledge or data beyond those we have considered will always be available to further inform ecosystem models; regardless, our results reveal that the choice of constraints can matter. 

Predicted population trajectories can appear qualitatively different based on the constraints (Figure \ref{fig:3_species_trajectories}). Population hindcasts and predictions in response to conservation actions produced by ensembles can yield populations that vastly different magnitudes and with different overall trends across constraint sets (Figure \ref{fig:dingo_decision}). These differences can be so significant that it may lead to different conservation decision-making based on the framework chosen for modelling. We found that the choice of constraints could lead to entirely different long-term behaviour (Figure \ref{fig:implied feas and stab}), such that population trajectories that appear reasonable according to our constraints are not capable of stable coexistence at equilibrium. This has the potential for dramatic impact on the management of complex ecosystems, so motivates careful consideration of the models that represent these ecosystems and the confidence in any forecasted scenarios in these ecosystems \citep{adams2020_TS,botelho_2024}.

\subsection{Robust statistical frameworks drive advancements in simulation models and their connection to new data sources}

Our novel framework connecting time series simulation models to expert knowledge in a statistically robust and efficient manner is useful beyond the specific examples discussed here. In this work, we exclusively demonstrated our methods on generalised Lotka-Volterra model structures because of its ubiquity in the field \citep{Dougoud_2018,baker_2019,adams2020_TS}; however, other simulation models can be used. Any model that produces population trajectories can be used within this framework, including stochastic representations \citep{ives_2003,black_2012_stochastic,saether_2002_stochastic,bjorkvoll_2012_stochastic,hostetler_2012_stochastic}, spatial models \citep{melbourne_2011_spatial,rayner_2007_spatial,briscoe_2019_spatial,fordham_2013_spatial} or individual/agent-based frameworks \citep{harfoot_2014,fedriani_2018,breckling_2005_IBM}. Additionally, in this work, we have emphasised that embedding expert-elicited information is invaluable where data is noisy, sparse or unavailable; but this statistical framework can use expert knowledge in conjunction with a myriad of other data sources, including any available time series data. 

As ecological modelling advances, statistical frameworks must not only be flexible to new data sources, but also efficient in the process that incorporates them. In switching from calculating equilibrium abundances to simulating population trajectories, the computational demands of generating ensembles can go from minutes to days (Table S4). Practically, relaxing equilibrium constraints for realistic ecosystem networks requires computational developments such as new statistical modifications to ensemble generation processes. In this work, we found that a temporally adaptive modification of the present state-of-the-art model generation algorithm \citep{vollert_2023_SMCEEM} could make the process up to four times faster (Table S4) without altering the results (Figure \ref{fig:algorithm comparison}). 

This temporal adaptation can enhance approximate Bayesian parameterisations of time series simulations in any context, and our novel framework can be generalised to incorporate expert knowledge beyond this ecological application. Modelling and simulation have become an integral part of understanding the world around us -- from molecular systems \citep{schlick_2010_molecular} to vast environmental domains \citep{Geary_2020} -- and new expert insights or data sources constantly enhance our capability to predict what will happen in these systems \citep{monsalve_2022_systems,ma_2017_systems}. Here, our desire to better represent ecosystems drives simultaneous improvement in computational techniques and statistical frameworks, benefiting ecology and quantitative science in general.

\section*{Data availability}
The MATLAB code needed to replicate the results presented here is freely and publicly available on FigShare (DOI: \url{10.6084/m9.figshare.25679550}).

\begingroup
\bibliographystyle{chicagoa}

\input{Refs.bbl}
\endgroup

\section*{Acknowledgements}
SAV is supported by a Queensland University of Technology Centre for Data Science, Australia Scholarship. CD is supported by an Australian Research Council Future Fellowship (FT210100260). MPA and SAV acknowledge funding support from an Australian Research Council Discovery Early Career Researcher Award (DE200100683). Computational resources were provided by the eResearch Office, Queensland University of Technology.

\section*{Author contributions}
SAV wrote the code and drafted the manuscript. SAV, CD and MPA designed the research, analysed the results, and edited the manuscript.

\newpage
\setcounter{subsection}{0}
\setcounter{table}{0}
\setcounter{figure}{0}
\setcounter{equation}{0}
\renewcommand{\thesubsection}{S.\arabic{subsection}}
\renewcommand{\thetable}{S\arabic{table}}
\renewcommand{\thefigure}{S\arabic{figure}}
\renewcommand{\theequation}{S\arabic{equation}}

{\LARGE \textbf{Supplementary materials}}

\subsection{Further mathematical detail for constraint sets}
\label{SM: math constraint sets}
Ecosystem models typically forecast multiple populations through time simultaneously. There are a multitude of suitable modelling frameworks that take the general form  
\begin{equation}
    \frac{\mathrm{d}n_i}{\mathrm{d}t}= f(n_i), \qquad i=1,\dots, N,
    \label{eq: model}
\end{equation}
where $\frac{\mathrm{d}n_i}{\mathrm{d}t}$ is the change in population abundances $n_i(t)$ for ecosystem node $i$ over time $t$, $N$ is the number of ecosystem nodes being modelled, and $f$ is the functional form of the model. This model is specified by a set of parameters, whose values are chosen such that the resulting model meets the ecosystem constraints (e.g.,\ feasible and stable equilibria). Here, we mathematically describe four sets of assumptions that could be used for constraining parameter values for modelling ecosystems populations.

\subsubsection{Positive and stable equilibria}
\label{SM: math positive and stable}

Feasibility and stability constrain the asymptotic (long-term) behaviour of the system. Hence, mathematically they are considered at the equilibrium population $n_i^*$ for each species $i$, obtained by solving 
\begin{equation}
    \frac{\mathrm{d}n_i^*}{\mathrm{d}t}= 0, \qquad i=1,\dots, N,
\end{equation}
for the model defined by Equation (1) in the manuscript. Feasibility is achieved if these equilibrium populations are positive, $n^*_i>0$ for all species $i$ \citep{Grilli_2017_FS}. 

Stability -- specifically local asymptotic stability -- deals with the behaviour of the system when it is near this equilibria, hence the Jacobian matrix must be evaluated at the equilibrium populations  $n_i^*$ to analyse stability: 
\begin{equation}
    J_{ij} = \left. \frac{\partial f_i}{\partial n_j}  \right| _{n_i=n_i^*},
\end{equation}
where $J_{ij}$ is the $(i,j)$th element of the Jacobian matrix, and $f_i$ is the change in abundance for the $i$th node represented by the RHS of Equation (1) in the manuscript. A system is stable if the ecosystem populations will return to the equilibrium populations \citep{Grilli_2017_FS}, such that the real part of all eigenvalues ($\lambda_i$) of the Jacobian matrix are negative, i.e. $\mathbb{R} \{ \lambda_i\}<0,  \ \forall i=1,...,N$. 

\subsubsection{Bounded and stable equilibria}
In addition to stability, the equilibrium population  $n_i^*$ of each species $i$ must be between $n_{i,\mathrm{min}}$ and $n_{i,\mathrm{max}}$ in its long-term dynamics (where $0 \leq n_{i,\mathrm{min}} < n_{i,\mathrm{max}}$). These bounds are specified for each species in the system where it is appropriate to do so, and for the remaining species setting $n_{i,\mathrm{min}}=0$ and $n_{i,\mathrm{max}}=\infty$ is equivalent to the feasibility condition.

\subsubsection{Bounded trajectories}
Where the system must be simulated for a period of time, the initial populations $\bm{n}(0)$ are required to calculate population trajectories. We use the previously specified population bounds $n_{i,\mathrm{min}}$ and $n_{i,\mathrm{max}}$ to define a prior distribution for initial populations $\pi(n_i(0)) \sim \mathcal{U}(n_{i,\mathrm{min}}, n_{i,\mathrm{max}})$ and these initial conditions $\pi(n_i(0))$ are treated as parameters to be calibrated. Mathematically, the constraints will be satisfied if $n_{i,\mathrm{min}} < n_i(t) < n_{i,\mathrm{max}}$ for all species $i$ and for the full period time $t = (0,T)$, where $n_0$ are the initial conditions. Since the equilibrium behaviour is no longer considered, stability of the system is disregarded. 

\subsubsection{Bounded and slow-changing trajectories}
The rate of change in species $i$ abundance per time step can be calculated as
\begin{equation}
    R_i(t+\delta t) = \frac{\mathrm{d}n_i(t)/\mathrm{d}t}{n_i(t)} = \frac{n_i(t+\delta t) - n_i(t)}{n_i(t) \delta t},
\end{equation}
where $n_i(t)$ is the population of species $i$ at time $t$, and $\delta t$ is the change in time. By defining and measuring this rate of change, constraints can be enforced to limit this rate of change such that $R_{i,\mathrm{decrease}} < R_i(t + \delta t) < R_{i,\mathrm{increase}}$, where $R_{i,\mathrm{decrease}}$ is the negative rate of change limiting how quickly populations can decrease, and $R_{i,\mathrm{increase}}$ is the positive rate of change limiting how quickly populations can increase. For example,\ setting $R_{i,\mathrm{decrease}} = -0.5$ would indicate populations cannot reduce to 50\% in one time step, and $R_{i,\mathrm{increase}} = 0.5$ would mean populations cannot increase by 50\% in one time step.) If it is inappropriate to limit the rate of change for a species $i$, the limits can be set as  $R_{i,\mathrm{decrease}} =-\infty$ and $R_{i,\mathrm{increase}}=\infty$, such that any fluctuations will be allowed. 

\subsubsection{Summary of mathematical detail for each set of constraints}
For each set of constraints, the mathematical description of the constraints and the measure used to calculate the discrepancy when using SMC-EEM are specified in Table \ref{SM tab: math constraint summary}. 

\begin{table}[H]
    \footnotesize
    \centering
    \begin{tabular}{p{0.15\linewidth} |p{0.2\linewidth} |p{0.2\linewidth}|p{0.3\linewidth} }
         \textbf{Set of constraints} & \textbf{Description} & \textbf{Mathematical definition} & \textbf{Discrepancy measure $\rho(\bm{\theta})$ used in SMC-EEM}  \\ \hline \hline
          (1) Positive and stable equilibria  
             & The long term behaviour of the system will have coexistence of all species, and will be able to recover from small changes to populations. The long term trajectories will be attracted towards a fixed set of positive populations, either reaching these populations or oscillating around them. 
             & Equilibrium populations are positive,  $ \ n_i^*>0 \ \ \forall i$, and stable, $\ \mathbb{R}(\lambda_i) <0\ \  \forall i$.  & { \begin{align*}
                 \sum_{i=1}^{N} \Bigl[&  \max \{ 0, -n^*_i(\bm{\theta}) \} \\
                 &+ \max\{0,\mathbb{R}\{ \lambda_i\}\} \Bigr]
             \end{align*} } \\ \hline
          (2) Bounded and stable equilibria 
             & The long term behaviour of the system will have reasonable populations of all species, and will be able to recover from small changes to populations. The long term trajectories will be attracted towards a fixed set of reasonable populations, either reaching these populations or oscillating around them.
             & Equilibrium populations are within bounds, $\ n_{i,\mathrm{min}} \leq n_i^* \leq n_{i,\mathrm{max}} \ \ \forall i$, and stable, $\ \mathbb{R}(\lambda_i)<0 \ \ \forall i$. & {\begin{align*}
                 \sum_{i=1}^{N} \Bigl[& \max \{ 0,  n^*_i(\bm{\theta}) - n_{i,\mathrm{max}} \} \\
                 &+ \max \{ 0,   n_{i,\mathrm{min}} - n^*_i(\bm{\theta}) \} \\
                 &+  \max\{0, \mathbb{R}\{ \lambda_i\} \} \Bigr]
             \end{align*}}\\ \hline
          (3) Bounded trajectories
             & For a specified period of time, the population sizes will be reasonable without any constraints on the population dynamics. Beyond this period and for different species abundances, populations are not limited.  
             & For $t=(0,T)$ populations are within bounds, $\ n_{i,\mathrm{min}} \leq n_i(t) \leq n_{i,\mathrm{max}} \ \ \forall i$. & {\begin{align*}
                \sum_{i=1}^{N} \int_{t=0}^T \Bigl[& \max \{ 0,   n_i(t,\bm{\theta})  - n_{i,\mathrm{max}} \} \mathrm{d}t  \\
                & + \max \{ 0,  n_{i,\mathrm{min}} - n_i(t,\bm{\theta}  \} \mathrm{d}t \Bigr]
             \end{align*}}\\ \hline
          (4) Bounded and slow-changing trajectories 
             & For a specified period of time, the population sizes will be reasonable and are limited by how rapidly they can increase or decrease. Beyond this period and for different species abundances, populations sizes and changes are not limited.
             & For $t=(0,T)$ populations are within bounds, $\ n_{i,\mathrm{min}} \leq n_i(t) \leq n_{i,\mathrm{max}} \ \ \forall i$ and the rate of population changes are within bounds, $\ R_{i,\mathrm{decrease}} \leq R_i(t) \leq R_{i,\mathrm{increase}} \ \ \forall i$. & {\begin{align*}
                 \sum_{i=1}^{N}  \int_{t=0}^T \Bigl[& \max \{ 0,   n_i(t,\bm{\theta})  - n_{i,\mathrm{max}} \} \mathrm{d}t \, \\
                 &+ \max \{ 0,  n_{i,\mathrm{min}} - n_i(t,\bm{\theta}  \} \mathrm{d}t \, \\
                 &+ \max\{0,  R_i(t,\bm{\theta})  -R_{i,\mathrm{increase}} \} \mathrm{d}t \, \\
                 &+ \max\{0,   R_{i,\mathrm{decrease}} - R_i(t,\bm{\theta})  \} \mathrm{d}t  \Bigr]
             \end{align*}}  
    \end{tabular}
    \caption{A detailed mathematical summary of each set of constraints.}
    \label{SM tab: math constraint summary}
\end{table}

\newpage
\subsection{Sequential Monte Carlo ensemble ecosystem modelling algorithms}
\label{SM:algorithms}

\citet{vollert_2023_SMCEEM} introduced the SMC-EEM algorithm, based on sequential Monte Carlo - approximate Bayesian computation and provides a general overview, detailed algorithm and code. Here, we detail the temporally adaptive version of the algorithm by highlighting the differences from SMC-EEM in {\color{blue} blue} via an overview (Algorithm \ref{Alg:overview of temporally adaptive SMC-EEM}) and in full (Algorithms \ref{Alg:temporally adaptive SMC-EEM} and \ref{Alg:temporally adaptive MCMC}). 

Within each iteration of the SMC-EEM algorithm (the black text in Algorithm \ref{Alg:overview of temporally adaptive SMC-EEM}), the ensemble of parameter sets are used to simulate the system (e.g.,\ produce population trajectories), such that each parameter set can be attributed a discrepancy: a measure of how much the constraints were broken (e.g., how much above the upper bound did populations go). Using the collection of discrepancy scores, new parameter sets are proposed to reduce the discrepancy of simulations and iteratively approach an ensemble which meets all constraints.

\begin{algorithm}[H]
\small
    \caption{Temporally adaptive SMC-EEM overview algorithm}
    \label{Alg:overview of temporally adaptive SMC-EEM}
    \textbf{INITIALISE} \\
    Define the discrepancy function as a measure of how much the constraints are exceeded, $\rho(\bm{\theta})$ \\
    Generate a sample from the prior distribution, $\pi(\bm{\theta})$
    \BlankLine
    \textbf{WEIGHT} \\
    Simulate the time series for each parameter set $\bm{\theta}$ {\color{blue}for the period $(0,t_0)$} \\
    Evaluate the discrepancy $\rho(\bm{\theta})$ for the simulation {\color{blue}from $(0,t_0)$}  \\   
    Set the initial discrepancy threshold $\epsilon_0$ such that a percentage of particles have $\rho<\epsilon_0$ 
    \BlankLine
    \While{there are particles that break the constraints, $\max(\bm{\rho}) > 0$ {\color{blue} \textbf{OR} the simulation time is less than the full period, $t_k<T$}} {
    \BlankLine
    {\color{blue} \If{there is a large portion of particles above the discrepancy threshold $\epsilon_t$ \textbf{OR} the simulation time is the full period $t_k=T$}{ \color{black}
    \textbf{RESAMPLE} \\
    Duplicate parameter sets based on their discrepancy ${\rho}$ to replace those with $\rho_i>\epsilon_t$ \\
    \BlankLine
    \textbf{MOVE} \\
    Move each replaced parameter set using MCMC-ABC}} 
    \BlankLine
    \textbf{REWEIGHT} \\
    {\color{blue} Increase the simulation time based on the percentage of $\rho=0$, such that $t_{k} = \min\{t_{k-1}(1+n_c/M), T\}$}
    Simulate the time series for each parameter set for the period {\color{blue}$(t_{k-1},t_k)$} \\
    Evaluate the discrepancy ${\rho}(\bm{\theta})$ for the simulation {\color{blue} from $(0,t_k)$}\\    
    Set the discrepancy threshold $\epsilon_t$, such that a percentage of particles have $\rho<\epsilon_t$}
\end{algorithm}

\begin{algorithm}[H]
\small
    \caption{Temporally adaptive SMC-EEM}
    \label{Alg:temporally adaptive SMC-EEM}
    \textbf{INITIALISE} \\
    Define the discrepancy function as a measure of how much the constraints are exceeded, $\rho(\bm{\theta})$ \\
    Specify the prior distribution, $\pi(\bm{\theta})$ \\
    Select the tuning variables, including: \\ 
    $\rightarrow$ The number of particles to be sampled, $M$ \\ 
    $\rightarrow$ The percentage of particles retained in each sequential step, $a$ \\
    $\rightarrow$ The desired probability of particles unmoved during MCMC-ABC, $c$ \\
    $\rightarrow$ The number of trial MCMC-ABC steps to gauge acceptance rate, $n_{\mathrm{MCMC}}$ \\
    {\color{blue} $\rightarrow$ The initial period to simulate, $t_k = t_{0}$ }\\
    {\color{blue} $\rightarrow$ The maximum number of particles that can be kept when resampling, $b$} \\
    Generate a sample of $M$ particles ($\{\bm{\theta}_i\}_{i=1}^{M}$) from the prior distribution, $\pi(\bm{\theta})$     
    \BlankLine
    \textbf{REWEIGHT} \\
    Simulate the systems defined by $\{\bm{\theta}_{i}\}_{i=1}^{M}$ for the period {\color{blue}$(0,t_k)$ } \\
    Evaluate the discrepancy $\bm{\rho} = \{\rho(\bm{\theta}_{i})\}_{i=1}^{M}$ for the simulation {\color{blue} from $(0,t_k)$} \\   
    Sort the particles $\bm{\theta}$ in ascending order of their corresponding discrepancy $\bm{\rho}$\\
    Set the discrepancy threshold $\epsilon_t$ based on the number of particles to be retained $n_{\mathrm{keep}}=\lfloor a\times M \rfloor$ 
    \BlankLine
    \While{there are particles that break the constraints, $\max(\bm{\rho}) > 0$ {\color{blue} \textbf{OR} the simulation time is less than the full period, $t_k<T$}} {
    \BlankLine
    {\color{blue} \If{there are enough particles to drop, $n_{keep}<b$ OR the simulation time is the full period $t_k=T$}{ \color{black}
    \textbf{RESAMPLE} \\
    Transform current values of $\bm{\theta}$ to $\tilde{\bm{\theta}}$ such that the parameter-space is less restricted \\
    Duplicate retained particle values based on discrepancy $\bm{\rho}$ to replace those with $\rho_i>\epsilon_t$ \\
    Calculate the sample covariance matrix, $\Sigma = cov(\{\bm{\theta}_{i}\}_{i=1}^{n_{\mathrm{keep}}})$ 
    \BlankLine
    \textbf{MOVE} \\
    \For{each of the $n_{\mathrm{MCMC}}$ trial MCMC-ABC steps}{
        Move the particles using MCMC-ABC (Algorithm \ref{Alg:temporally adaptive MCMC})
    }
    Estimate the acceptance rate, $a_t$ \\
    Determine the number of MCMC-ABC iterations to perform, $R_t = \lceil \log (c)/\log(1-a_t) \rceil$ \\
    \For{each of the remaining MCMC-ABC steps, $R_t-n_{\mathrm{MCMC}}$}{
        Move the particles using MCMC-ABC (Algorithm \ref{Alg:temporally adaptive MCMC})
    }}}
    \BlankLine
    \textbf{REWEIGHT} \\
    {\color{blue} Increase the simulation time based on the number of particles $n_c$ with no discrepancy, such that $t_{k} = \min\{t_{k-1}(1+n_c/M), T\}$}} 
    Transform current values of $\tilde{\bm{\theta}}$ to $\bm{\theta}$ to simulate the system \\
    Simulate the systems defined by $\{\bm{\theta}_{i}\}_{i=1}^{M}$ for the period {\color{blue}$(t_{k-1},t_k)$ (NOTE: $(0,t_{k-1})$ is already simulated) } \\
    Evaluate the discrepancy $\bm{\rho} = \{\rho(\bm{\theta}_{i})\}_{i=1}^{M}$ for the simulation {\color{blue} from $(0,t_k)$}\\    
    Sort the particles $\bm{\theta}$ in ascending order of their corresponding discrepancy $\bm{\rho}$\\
    Set the discrepancy threshold based on the number of particles to be retained, $\epsilon_t = \rho(\bm{\theta}_{n_{\mathrm{keep}}})$\\
    \If{we are dropping particles which meet all constraints, $\rho(\bm{\theta}_i) =0$ where $i > n_{\mathrm{keep}}$}{
    Adjust $n_{\mathrm{keep}}$ to retain all particles where $\rho(\bm{\theta}_i) =0$
    }
\end{algorithm}

\begin{algorithm}[H]
\small
        \For{each particle $i$ in $\{\bm{\theta}_{i}\}_{i=n_{\mathrm{keep}}}^{M}$}{
        Propose a new set of parameter values  ${\tilde{\bm{\theta}_i}}^*$ using a multivariate normal proposal distribution, ${\tilde{\bm{\theta}_i}}^* \sim N(\tilde{\bm{\theta}_i},\Sigma)$ \\        
        Calculate the prior probability ($\pi(\tilde{\bm{\theta}})$ for the transform space) of the current and proposed parameter values ($\tilde{\bm{\theta}_i}$ and ${\tilde{\bm{\theta}_i}}^*$) \\
        Transform the current and proposed parameter values ($\tilde{\bm{\theta}_i}$ and ${\tilde{\bm{\theta}_i}}^*$) in terms of $\bm{\theta}$ \\
        Simulate the system defined by $\{\bm{\theta}_{i}^*\}$ for the period, {\color{blue} $(0,t_k)$}  \\
        Evaluate the discrepancy $\rho(\bm{\theta}_{i}^*)$ for the simulation {\color{blue} from $(0,t_k)$}\\
        Accept or reject a particle based on a Metropolis-Hastings acceptance probability $\alpha = \min \left(1, \pi(\tilde{\bm{\theta}}_i^*)/\pi(\tilde{\bm{\theta}}_i) \right)$, if within the discrepancy threshold, $\rho(\bm{\theta}_{i}^*) \leq \epsilon_t$ 
        }
    \caption{MCMC-ABC algorithm used within temporally adaptive SMC-EEM method (Algorithm \ref{Alg:temporally adaptive SMC-EEM}) }
\label{Alg:temporally adaptive MCMC}
\end{algorithm}

\newpage
\subsection{Additional information for the three species predator-prey network}
This section contains additional details of the three-species ecosystem model (Section \ref{SM:3sp_details}) and comparisons between models produced using different sets of constraints (Section \ref{3sp constraints comparison figs}). 

\subsubsection{Modelling details}
\label{SM:3sp_details}
The three-species ecosystem network (depicted in Figure 1a in the main text) was modelled using the generalised Lotka-Volterra equations. The generalised Lotka-Volterra equations take the form
\begin{equation}
    \frac{\mathrm{d}n_i}{\mathrm{d}t}= \left[ r_i + \sum_{j=1}^{N} \alpha_{i,j} n_j(t) \right] n_i(t),
    \label{eq: LV}
\end{equation}
where $n_i(t)$ is the abundance of the $i$th ecosystem node at time $t$, $r_i$ is the growth rate of the $i$th ecosystem node, $N$ is the number of ecosystem nodes being modelled, and $\alpha_{i,j}$ is the per-capita interaction strength characterising the effect of node $j$ on node $i$. Typically, an ecosystem network -- which specifies the interactions between each species -- are used to characterise whether each interaction is positive, negative or zero. For this three-species network, the associated ecosystem model is given by 
\begin{align*}
    \frac{\mathrm{d}n_A}{\mathrm{d}t} &= r_A n_A - \alpha_{A,A} {n_A}^2 +  \alpha_{A,B} n_B n_A, \\
    \frac{\mathrm{d}n_B}{\mathrm{d}t} &= r_B n_B - \alpha_{B,A} n_A n_B -  \alpha_{B,B} {n_B}^2 + \alpha_{B,C} n_B n_C, \\
    \frac{\mathrm{d}n_C}{\mathrm{d}t} &= r_C n_C -  \alpha_{C,B} {n_C} n_B - \alpha_{C,C} {n_C}^2 
\end{align*}

The modelling framework used to generate the results was the Lotka-Volterra equations, though any model structure which allows calculation of the equilibrium populations and stability can be used instead. The equilibrium population $n_i^*$ for each species $i$ is the solution to  
\begin{equation}
    \frac{\mathrm{d}n^*_i}{\mathrm{d}t}=r_i n^*_i + n^*_i \sum_{j=1}^{N} \alpha_{i,j} n^*_j = 0, \quad \quad \quad \forall i=1,...,N.
    \label{eq: equilibrium}
\end{equation}
For this three-species model, this can be calculated as ...

To determine if the equilibrium is stable first requires calculation of the Jacobian matrix, evaluated at equilibrium populations ${n}_i^*$, 
\begin{equation}
    J_{ij} = \left. \frac{\partial \left(  r_i n_i + n_i \sum_{j=1}^{N} \alpha_{i,j} n_j \right)}{\partial n_j}  \right| _{n_i=n_i^*}, 
    \label{eq:jacobian}
\end{equation}
where $J_{ij}$ is the $(i,j)$th element of the Jacobian matrix. The dynamic system is considered locally asymptotically stable if the real part of all eigenvalues ($\lambda_i$) of the Jacobian matrix are negative, i.e. $\mathbb{R} \{ \lambda_i\}<0,  \ \forall i=1,...,N$. 

Within this modelling framework, the following details were specified: 
\begin{table}[H]
    \centering
    \begin{tabular}{m{0.2\textwidth}|m{0.35\textwidth}|m{0.25\textwidth}}
        \textbf{Item} & \textbf{Description} & \textbf{Choice}  \\ \hline \hline
        Ensemble size & The number of parameter sets to be generated that meet the constraints. & $M = 10,000$\\ \hline
        Prior distribution & A distribution of the initial belief of parameter values. & $r_i \sim U(-5,5),$  \\ & & $\alpha_{i,j} \sim U(0,1), $\\&&$\qquad  \forall i,j = {A,B,C}.$ \\ \hline
        Population bounds  & The (hypothetical) expert-elicited population limits for each species. & $5 \leq n_{\mathrm{A}} \leq 15, $ \\ & &$0 \leq n_{\mathrm{B}} \leq 25, $ \\ & &$0 \leq n_{\mathrm{C}} \leq 30. $\\ \hline
        Simulation period & The observation period for trajectory-based constraints. & $T = 1$ \\ \hline
        Population change bounds & The (hypothetical) expert-elicited limits on population rate of change & $ -1 \leq R_A \leq 1, $ \\ & & $-2 \leq R_B \leq 2,$ \\ & & $-6 \leq R_C \leq 6.$
    \end{tabular}
    \caption{Modelling choices for the three-species predator-prey ecosystem case study.  }
    \label{tab:3sp modelling details}
\end{table}

\subsubsection{Additional results comparing sets of constraints}
\label{3sp constraints comparison figs}
\begin{figure}[H]
    \centering
    \includegraphics[width=\textwidth]{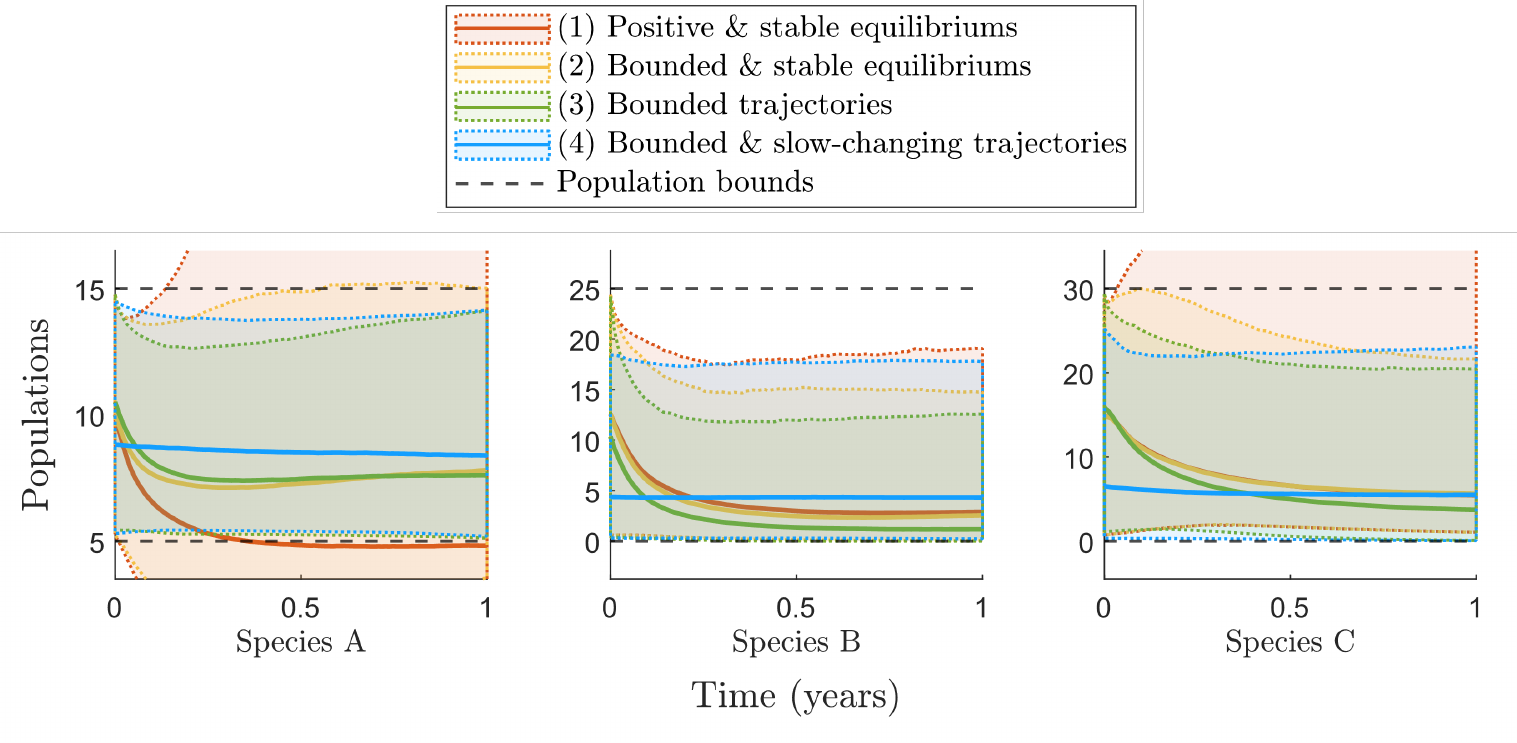}
    \caption{Median and 95\% prediction interval populations for the three-species ecosystem model using each of the four sets of constraints. }
    \label{fig:3sp_credible_intervals}
\end{figure}

\begin{figure}[H]
    \centering
    \includegraphics[width=0.8\textwidth]{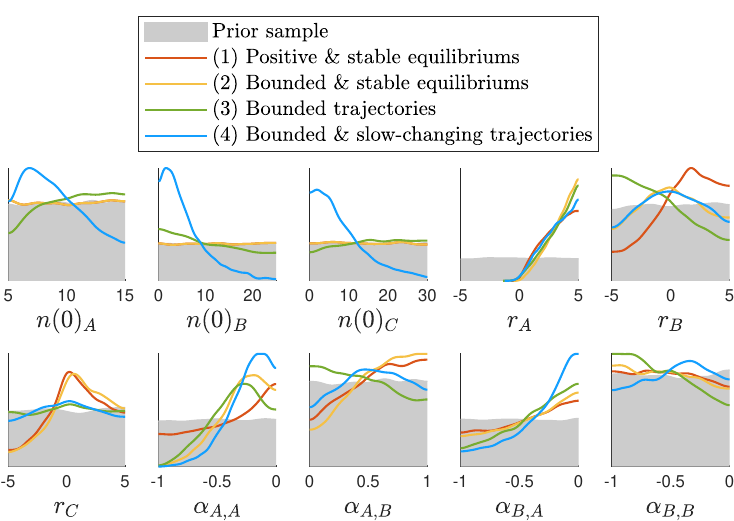}
    \caption{Estimated marginal parameter distributions of the three-species predator-prey ecosystem model when parameterised to each of the four sets of constraints. }
    \label{fig:3sp_parameter_dists}
\end{figure}

\newpage
\subsection{Additional information for the semiarid Australia case study}
\label{SM: dingo ALL}
This section contains additional detail on the modelling of the semiarid Australia ecosystem model (Section \ref{dingo details}), additional comparison of the effects of different sets of constraints on the modelling (Section \ref{dingo constraints}), and additional results comparing SMC-EEM with temporally adaptive SMC-EEM (Section \ref{SM:Din_compare_algos}). 

\subsubsection{Modelling details}
\label{dingo details}
The ecosystem model we consider for this example is the generalised Lotka Volterra model for the semiarid Australian ecosystem network (Figure 2a in the manuscript). For more details on the generalised Lotka Volterra equations and ecosystem networks see Supplementary Materials Section \ref{SM:3sp_details} where we describe the process for the three-species predator-prey network.  

Within this modelling framework, the following details were specified: 
\begin{table}[H]
    \centering
    \begin{tabular}{m{0.2\textwidth}|m{0.35\textwidth}|m{0.25\textwidth}}
        \textbf{Item} & \textbf{Description} & \textbf{Choice}  \\ \hline \hline
        Ensemble size & The number of parameter sets to be generated that meet the constraints. & $M = 50,000$\\ \hline
        Prior distribution & A distribution of the initial belief of parameter values. & $r_i \sim U(-5,5),  \ \ \forall i$  \\ & & $\alpha_{i,j} \sim U(0,1), \ \ \forall i,j.$ \\ \hline
        Population bounds  & The (hypothetical) expert-elicited population limits for each species. & $10 \leq n_{\mathrm{D}} \leq 100, $ \\ & &$10 \leq n_{\mathrm{M}} \leq 500, $ \\ & &$10 \leq n_{\mathrm{LH}} \leq 600, $ \\ & &$50 \leq n_{\mathrm{SV}} \leq 1000, $ \\ & &$100 \leq n_{\mathrm{G}} \leq 1000, $ \\ & &$100 \leq n_{\mathrm{I}} \leq 1000, $ \\ & &$0 \leq n_{\mathrm{F}} \leq 10, $ \\ & &$0 \leq n_{\mathrm{S}} \leq 1000$\\ \hline
        Simulation period & The observation period for trajectory-based constraints. & $T = 1$ \\ \hline
        Population change bounds & The (hypothetical) expert-elicited limits on population rate of change & $R_{i,\mathrm{increase}} = 12,  \ \  \forall i$ \\ & &$ R_{i,\mathrm{decrease}} = -36, \ \  \forall i.$
    \end{tabular}
    \caption{Modelling choices for the semiarid Australian ecosystem network case study. Population limits for each ecosystem network node are indicated such that $D$ represents dingoes, $M$ represents mesopredators, $LH$ represents large herbivores, $SV$ represents small vertebrates, $G$ represents grasses, $I$ represents invertebrates, $F$ represents fires, and $S$ represents soil quality. }
    \label{tab:dingo modelling details}
\end{table}

Using the ensembles parameterised using each set of constraints, we simulated the system for the observation period ($T=1$ year) then removed 50\% of the dingo population, and simulated the resulting populations over the following six months ($T=1/2$ years). 

\subsubsection{Additional results comparing sets of constraints}
\label{dingo constraints}
\begin{figure}[H]
    \centering
    \includegraphics[width=\textwidth]{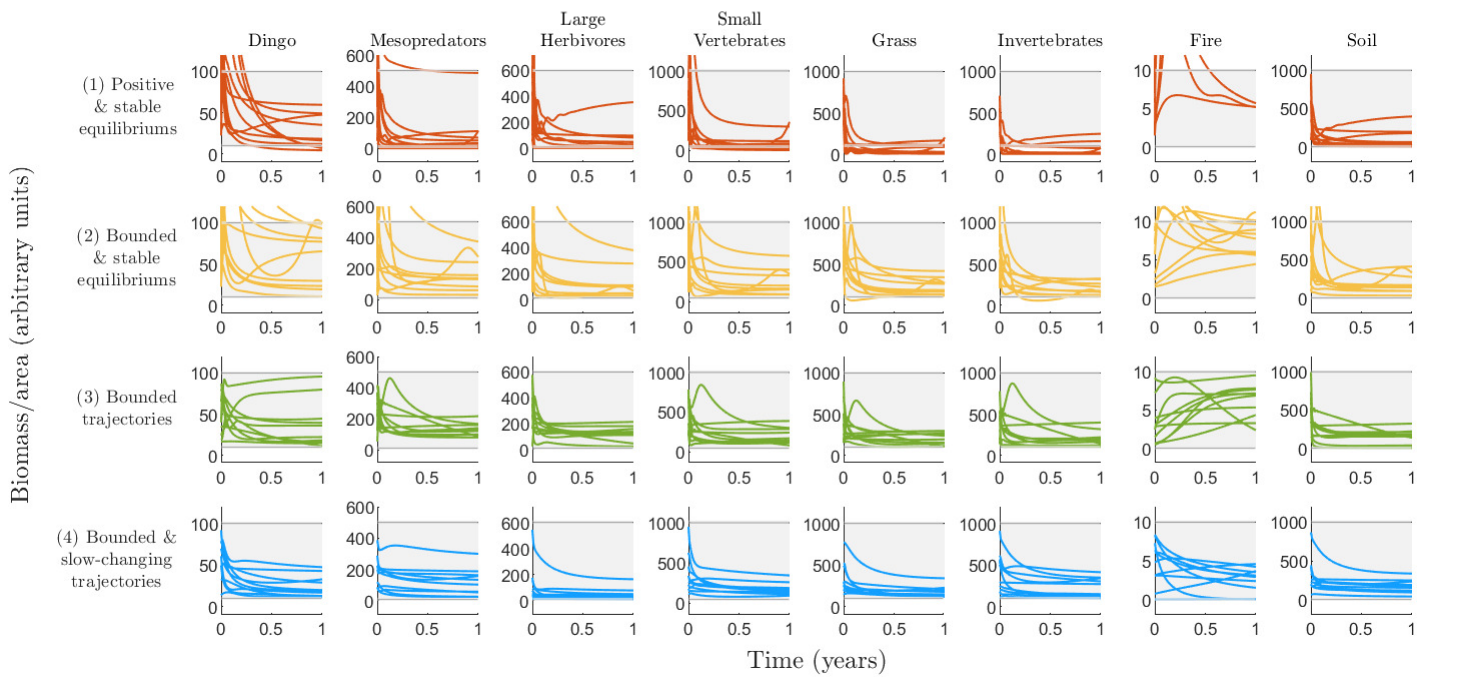}
    \caption{Comparison of some example trajectories produced by each set of constraints for the semiarid Australia ecosystem model. }
    \label{fig:dingo example trajectories}
\end{figure}

\begin{figure}[H]
    \centering
    \includegraphics[width=\textwidth]{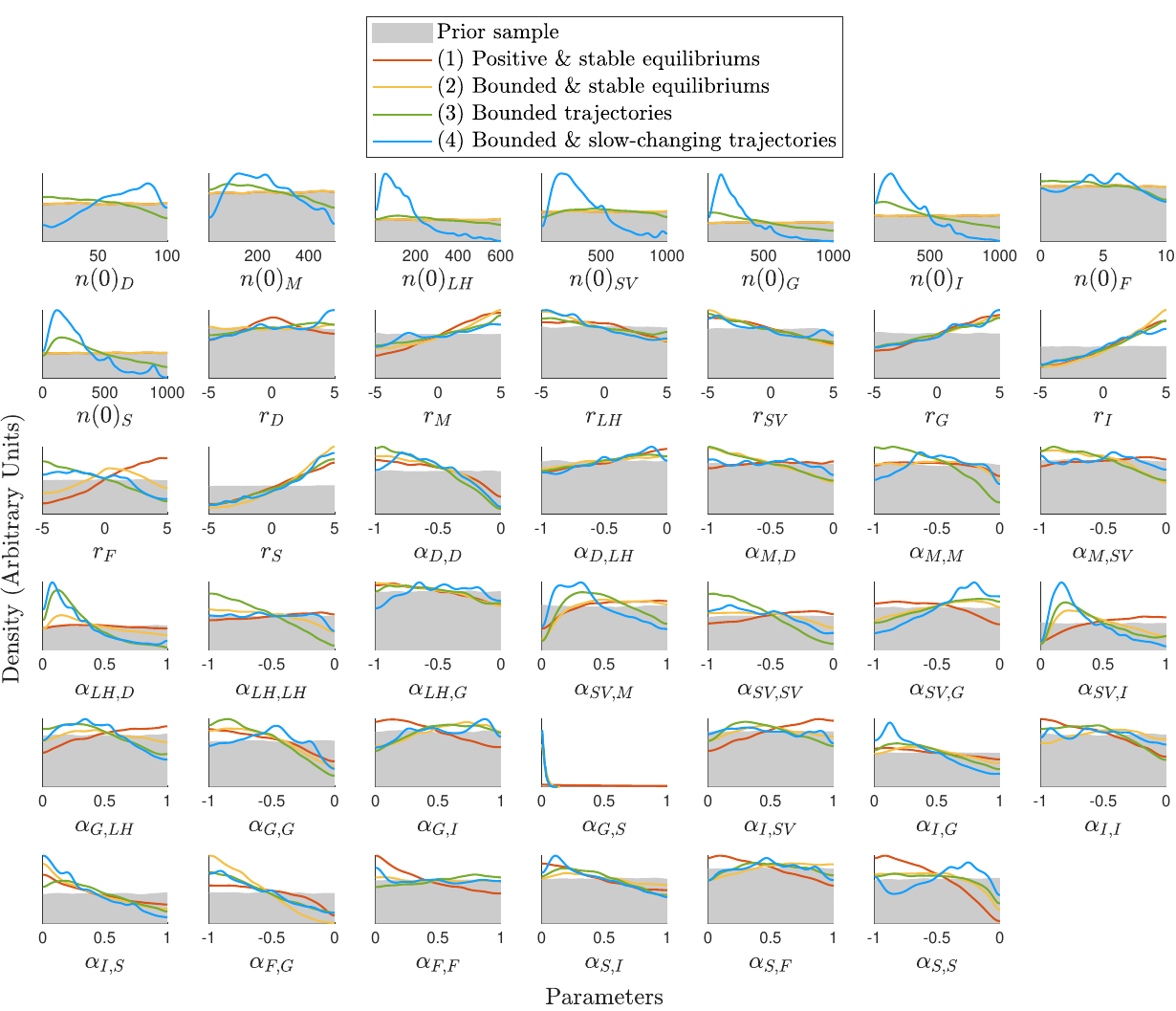}
    \caption{Estimated marginal parameter values for each set of constraints for the semiarid Australian ecosystem model. }
    \label{fig:dingo parameters}
\end{figure}

\subsubsection{Additional results comparing ensemble generation algorithms}
\label{SM:Din_compare_algos}
The two different sampling algorithms -- SMC-EEM and temporally adaptive SMC-EEM -- were compared to assess whether the outputs of SMC-EEM were reproducible using temporally adaptive SMC-EEM. In addition to the comparison of predictive distributions (Figure 4b), we can compare the estimated  marginal parameter values between SMC-EEM and the temporally adaptive version (see \citet{vollert_2023_SMCEEM} for a similar comparison).

\begin{figure}[H]
    \centering
    \includegraphics[width=\textwidth]{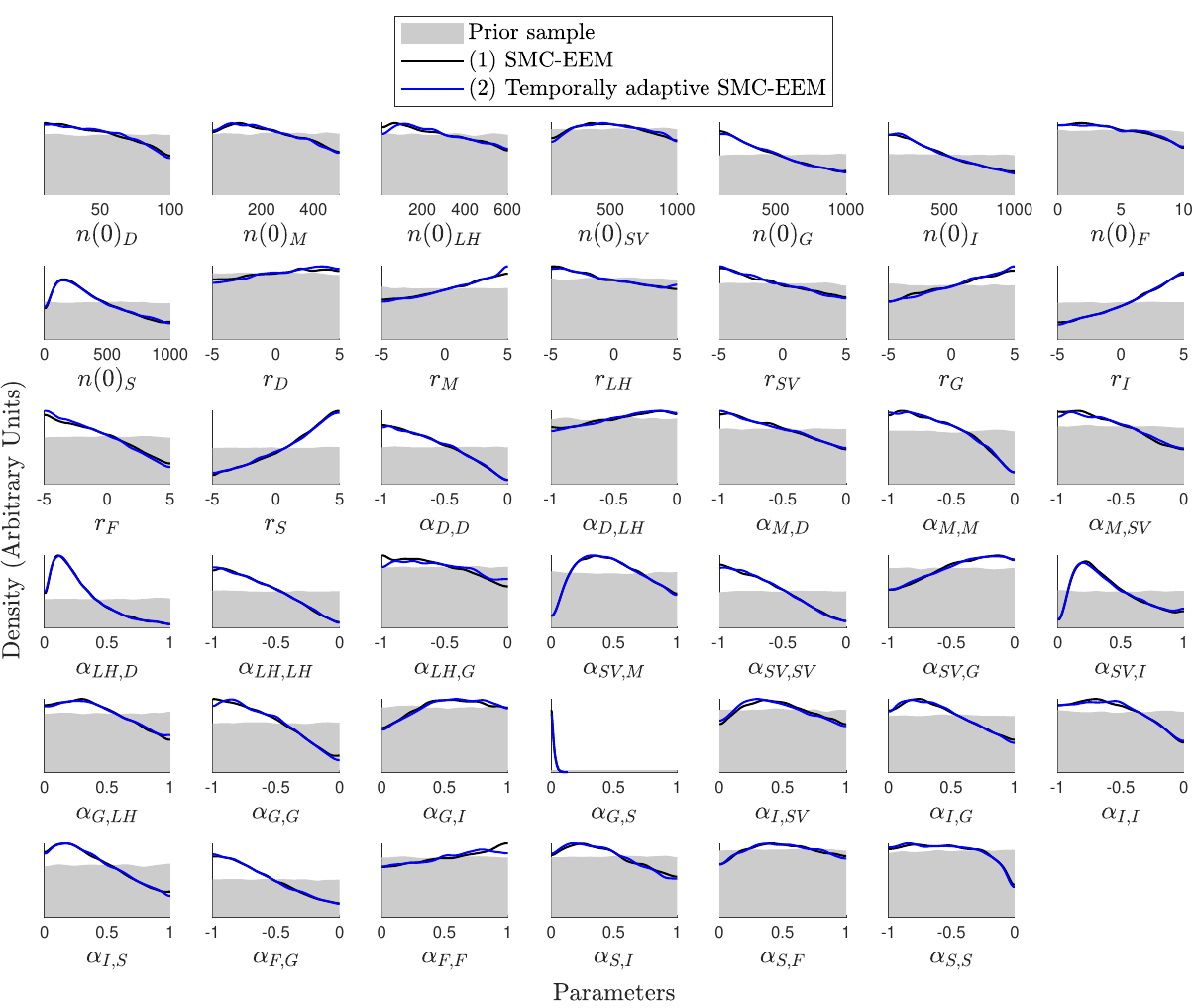}
    \caption{The estimated marginal parameter distributions for the semiarid Australia model, constrained using bounded trajectories (set 3). Here the outputs of SMC-EEM and temporally adaptive SMC-EEM are equivalent. }
    \label{fig:Dingrowth_params}
\end{figure}

\begin{table}[H]
    \centering
    \begin{tabular}{m{0.32\textwidth}|m{0.25\textwidth}|m{0.25\textwidth}}
        \textbf{Set of constraints} & \multicolumn{2}{c}{\textbf{Computation time (seconds)}} \\
                                    & \textbf{SMC-EEM} & \textbf{Temporally adaptive SMC-EEM} \\ \hline
         (1) Positive \& stable equilibria & $70$ & NA \\
         (2) Bounded \& stable equilibria & $1086$ ($\sim 18.1$ minutes) & NA\\
         (3) Bounded trajectories & $12232$ ($\sim 3.4$ hours) & $6675$ ($\sim 1.9$ hours) \\
         (4) Bounded \& slow-changing trajectories & $294582$ ($\sim 81.8$ hours) & $63841$ ($\sim 17.7$ hours) \\
    \end{tabular}
    \caption{Computation times for generating 50000 parameter sets the semiarid Australia ecosystem model with each set of constraints. These ensembles were generated using the Queensland University of Technology's high-performance computers with 12 parallel cores. }
    \label{tab:dingo_times}
\end{table}

\end{document}

%% file: titlepage.tex
\begin{titlepage}
    
    \begin{center}
   
    \vspace*{2cm}
    \LARGE
    \textbf{Ecosystem knowledge should replace coexistence and stability assumptions in ecological network modelling}
       
    \vspace{1cm}
    \large{Sarah A. Vollert$^{a,b,*}$, Christopher Drovandi$^{a,b}$, \& Matthew P. Adams$^{a,b,c}$} \\

    \vspace{1cm}
    \normalsize
    $^a$Centre for Data Science, Queensland University of Technology, Brisbane, Australia \\

    $^b$School of Mathematical Sciences, Queensland University of Technology, Brisbane, Australia\\

    $^c$School of Chemical Engineering, The University of Queensland, St Lucia, Australia\\

    $^*$Corresponding author. E-mail: sarah.vollert@hdr.qut.edu.au
  
   \end{center}

\newpage
{\large \textbf{Abstract}} 

\begin{enumerate}
    \item Quantitative population modelling is an invaluable tool for identifying the cascading effects of ecosystem management and interventions. Ecosystem models are often constructed by assuming stability and coexistence in ecological communities as a proxy for abundance data when monitoring programs are not available. However, a growing body of literature suggests that these assumptions are inappropriate for modelling conservation outcomes. In this work, we develop an alternative for dataless population modelling that instead relies on expert-elicited knowledge of species abundances. 
    \item While time series abundance data is often not available for ecosystems of interest, these systems may still be highly studied or observed in an informal capacity. In particular, limits on population sizes and their capacity to rapidly change during an observation period can be reasonably elicited for a number of species. This ecosystem knowledge may be a more trustworthy source of information for constructing models for conservation than theoretical assumptions of coexistence and stability. 
    \item We propose a robust framework for generating an ensemble of ecosystem models whose population predictions match the expected population dynamics, as defined by experts. Our new Bayesian algorithm systematically removes model parameters that lead to unreasonable population predictions without incurring excessive computational costs. We demonstrate our framework on a series of increasingly pragmatic constraints for population dynamics by incorporating expert-defined limits on population sizes and fluctuations and removing equilibrium requirements.
    \item Our results demonstrate that models constructed using expert-elicited information, rather than stability and coexistence assumptions, can dramatically impact population predictions, expected responses to management, conservation decision-making, and long-term ecosystem behaviour. In the absence of data, we argue that field observations and expert knowledge are preferred for representing ecosystems observed in nature instead of theoretical assumptions of coexistence and stability.
\end{enumerate}

{\large \textbf{Keywords}}\\ Approximate Bayesian computation; Coexistence; Community ecology; Conservation planning; Ensemble ecosystem modeling; Population modeling; Sequential Monte Carlo; Stability; 

\end{titlepage}

%% file: main.bbl
\begin{thebibliography}{}

\bibitem[\protect\citeauthoryear{Adams, Sisson, Helmstedt, Baker, Holden, Plein, Holloway, Mengersen, and McDonald-Madden}{Adams et~al.}{2020}]{adams2020_TS}
Adams, M.~P., S.~A. Sisson, K.~J. Helmstedt, C.~M. Baker, M.~H. Holden, M.~Plein, J.~Holloway, K.~L. Mengersen, and E.~McDonald-Madden (2020).
\newblock Informing management decisions for ecological networks, using dynamic models calibrated to noisy time-series data.
\newblock {\em Ecology Letters\/}~{\em 23\/}(4), 607--619.


\bibitem[\protect\citeauthoryear{Allesina and Tang}{Allesina and Tang}{2012}]{Allesina_2012_stab}
Allesina, S. and S.~Tang (2012).
\newblock Stability criteria for complex ecosystems.
\newblock {\em Nature\/}~{\em 483\/}(7388), 205--208.


\bibitem[\protect\citeauthoryear{Baker, Bode, Dexter, Lindenmayer, Foster, MacGregor, Plein, and McDonald-Madden}{Baker et~al.}{2019}]{baker_2019}
Baker, C.~M., M.~Bode, N.~Dexter, D.~B. Lindenmayer, C.~Foster, C.~MacGregor, M.~Plein, and E.~McDonald-Madden (2019).
\newblock A novel approach to assessing the ecosystem-wide impacts of reintroductions.
\newblock {\em Ecological Applications\/}~{\em 29\/}(1), e01811.


\bibitem[\protect\citeauthoryear{Baker, Gordon, and Bode}{Baker et~al.}{2017}]{baker2017_EEM}
Baker, C.~M., A.~Gordon, and M.~Bode (2017).
\newblock Ensemble ecosystem modeling for predicting ecosystem response to predator reintroduction.
\newblock {\em Conservation Biology\/}~{\em 31\/}(2), 376--384.


\bibitem[\protect\citeauthoryear{Bj{\o}rkvoll, Gr{\o}tan, Aanes, S{\ae}ther, Engen, and Aanes}{Bj{\o}rkvoll et~al.}{2012}]{bjorkvoll_2012_stochastic}
Bj{\o}rkvoll, E., V.~Gr{\o}tan, S.~Aanes, B.-E. S{\ae}ther, S.~Engen, and R.~Aanes (2012).
\newblock Stochastic population dynamics and life-history variation in marine fish species.
\newblock {\em The American Naturalist\/}~{\em 180\/}(3), 372--387.


\bibitem[\protect\citeauthoryear{Black and McKane}{Black and McKane}{2012}]{black_2012_stochastic}
Black, A.~J. and A.~J. McKane (2012).
\newblock Stochastic formulation of ecological models and their applications.
\newblock {\em Trends in ecology \& evolution\/}~{\em 27\/}(6), 337--345.


\bibitem[\protect\citeauthoryear{Bode, Baker, Benshemesh, Burnard, Rumpff, Hauser, Lahoz-Monfort, and Wintle}{Bode et~al.}{2017}]{Bode_2017}
Bode, M., C.~M. Baker, J.~Benshemesh, T.~Burnard, L.~Rumpff, C.~E. Hauser, J.~J. Lahoz-Monfort, and B.~A. Wintle (2017).
\newblock Revealing beliefs: using ensemble ecosystem modelling to extrapolate expert beliefs to novel ecological scenarios.
\newblock {\em Methods in Ecology and Evolution\/}~{\em 8\/}(8), 1012--1021.


\bibitem[\protect\citeauthoryear{Boettiger}{Boettiger}{2021}]{boettiger_2021}
Boettiger, C. (2021).
\newblock Ecological management of stochastic systems with long transients.
\newblock {\em Theoretical Ecology\/}~{\em 14\/}(4), 663--671.


\bibitem[\protect\citeauthoryear{Botelho, Jeynes-Smith, Vollert, and Bode}{Botelho et~al.}{2024}]{botelho_2024}
Botelho, L.~L., C.~Jeynes-Smith, S.~Vollert, and M.~Bode (2024).
\newblock Ecosystem models cannot predict the consequences of conservation decisions.
\newblock {\em arXiv preprint arXiv:2401.10439\/}.


\bibitem[\protect\citeauthoryear{Botkin}{Botkin}{1990}]{botkin_1990}
Botkin, D.~B. (1990).
\newblock Discordant harmonies: a new ecology for the twenty-first century.


\bibitem[\protect\citeauthoryear{Breckling, M{\"u}ller, Reuter, H{\"o}lker, and Fr{\"a}nzle}{Breckling et~al.}{2005}]{breckling_2005_IBM}
Breckling, B., F.~M{\"u}ller, H.~Reuter, F.~H{\"o}lker, and O.~Fr{\"a}nzle (2005).
\newblock Emergent properties in individual-based ecological models—introducing case studies in an ecosystem research context.
\newblock {\em Ecological modelling\/}~{\em 186\/}(4), 376--388.


\bibitem[\protect\citeauthoryear{Briscoe, Elith, Salguero-G{\'o}mez, Lahoz-Monfort, Camac, Giljohann, Holden, Hradsky, Kearney, McMahon, et~al.}{Briscoe et~al.}{2019}]{briscoe_2019_spatial}
Briscoe, N.~J., J.~Elith, R.~Salguero-G{\'o}mez, J.~J. Lahoz-Monfort, J.~S. Camac, K.~M. Giljohann, M.~H. Holden, B.~A. Hradsky, M.~R. Kearney, S.~M. McMahon, et~al. (2019).
\newblock Forecasting species range dynamics with process-explicit models: matching methods to applications.
\newblock {\em Ecology Letters\/}~{\em 22\/}(11), 1940--1956.


\bibitem[\protect\citeauthoryear{Broughton, Bullock, George, Hill, Hinsley, Maziarz, Melin, Mountford, Sparks, and Pywell}{Broughton et~al.}{2021}]{broughton_2021_woodland}
Broughton, R.~K., J.~M. Bullock, C.~George, R.~A. Hill, S.~A. Hinsley, M.~Maziarz, M.~Melin, J.~O. Mountford, T.~H. Sparks, and R.~F. Pywell (2021).
\newblock Long-term woodland restoration on lowland farmland through passive rewilding.
\newblock {\em PloS one\/}~{\em 16\/}(6), e0252466.


\bibitem[\protect\citeauthoryear{Cuddington}{Cuddington}{2001}]{Cuddington_2001}
Cuddington, K. (2001).
\newblock The “balance of nature” metaphor and equilibrium in population ecology.
\newblock {\em Biology and Philosophy\/}~{\em 16}, 463--479.


\bibitem[\protect\citeauthoryear{DeAngelis and Waterhouse}{DeAngelis and Waterhouse}{1987}]{deangelis_1987}
DeAngelis, D.~L. and J.~Waterhouse (1987).
\newblock Equilibrium and nonequilibrium concepts in ecological models.
\newblock {\em Ecological Monographs\/}~{\em 57\/}(1), 1--21.


\bibitem[\protect\citeauthoryear{Dexter, Ramsey, MacGregor, and Lindenmayer}{Dexter et~al.}{2012}]{dexter_2012}
Dexter, N., D.~S. Ramsey, C.~MacGregor, and D.~Lindenmayer (2012).
\newblock Predicting ecosystem wide impacts of wallaby management using a fuzzy cognitive map.
\newblock {\em Ecosystems\/}~{\em 15}, 1363--1379.


\bibitem[\protect\citeauthoryear{De’Ath, Fabricius, Sweatman, and Puotinen}{De’Ath et~al.}{2012}]{death_2012}
De’Ath, G., K.~E. Fabricius, H.~Sweatman, and M.~Puotinen (2012).
\newblock The 27--year decline of coral cover on the great barrier reef and its causes.
\newblock {\em Proceedings of the National Academy of Sciences\/}~{\em 109\/}(44), 17995--17999.


\bibitem[\protect\citeauthoryear{Donohue, Hillebrand, Montoya, Petchey, Pimm, Fowler, Healy, Jackson, Lurgi, McClean, et~al.}{Donohue et~al.}{2016}]{donohue_2016_stability}
Donohue, I., H.~Hillebrand, J.~M. Montoya, O.~L. Petchey, S.~L. Pimm, M.~S. Fowler, K.~Healy, A.~L. Jackson, M.~Lurgi, D.~McClean, et~al. (2016).
\newblock Navigating the complexity of ecological stability.
\newblock {\em Ecology Letters\/}~{\em 19\/}(9), 1172--1185.


\bibitem[\protect\citeauthoryear{Dougoud, Vinckenbosch, Rohr, Bersier, and Mazza}{Dougoud et~al.}{2018}]{Dougoud_2018}
Dougoud, M., L.~Vinckenbosch, R.~P. Rohr, L.-F. Bersier, and C.~Mazza (2018).
\newblock The feasibility of equilibria in large ecosystems: {A} primary but neglected concept in the complexity-stability debate.
\newblock {\em PLoS Computational Biology\/}~{\em 14\/}(2), e1005988.


\bibitem[\protect\citeauthoryear{Edelstein-Keshet}{Edelstein-Keshet}{2005}]{Edelstein-Keshet_2005_mathbio}
Edelstein-Keshet, L. (2005).
\newblock {\em Mathematical Models in Biology}.
\newblock SIAM.


\bibitem[\protect\citeauthoryear{Fedriani, Wiegand, Ayll{\'o}n, Palomares, Su{\'a}rez-Esteban, and Grimm}{Fedriani et~al.}{2018}]{fedriani_2018}
Fedriani, J.~M., T.~Wiegand, D.~Ayll{\'o}n, F.~Palomares, A.~Su{\'a}rez-Esteban, and V.~Grimm (2018).
\newblock Assisting seed dispersers to restore oldfields: an individual-based model of the interactions among badgers, foxes and iberian pear trees.
\newblock {\em Journal of applied ecology\/}~{\em 55\/}(2), 600--611.


\bibitem[\protect\citeauthoryear{Fordham, Ak{\c{c}}akaya, Brook, Rodr{\'\i}guez, Alves, Civantos, Trivi{\~n}o, Watts, and Ara{\'u}jo}{Fordham et~al.}{2013}]{fordham_2013_spatial}
Fordham, D.~A., H.~Ak{\c{c}}akaya, B.~W. Brook, A.~Rodr{\'\i}guez, P.~C. Alves, E.~Civantos, M.~Trivi{\~n}o, M.~J. Watts, and M.~B. Ara{\'u}jo (2013).
\newblock Adapted conservation measures are required to save the iberian lynx in a changing climate.
\newblock {\em Nature Climate Change\/}~{\em 3\/}(10), 899--903.


\bibitem[\protect\citeauthoryear{Francis, Abbott, Cuddington, Gellner, Hastings, Lai, Morozov, Petrovskii, and Zeeman}{Francis et~al.}{2021}]{francis_2021}
Francis, T.~B., K.~C. Abbott, K.~Cuddington, G.~Gellner, A.~Hastings, Y.-C. Lai, A.~Morozov, S.~Petrovskii, and M.~L. Zeeman (2021).
\newblock Management implications of long transients in ecological systems.
\newblock {\em Nature Ecology \& Evolution\/}~{\em 5\/}(3), 285--294.


\bibitem[\protect\citeauthoryear{Geary, Bode, Doherty, Fulton, Nimmo, Tulloch, Tulloch, and Ritchie}{Geary et~al.}{2020}]{Geary_2020}
Geary, W.~L., M.~Bode, T.~S. Doherty, E.~A. Fulton, D.~G. Nimmo, A.~I. Tulloch, V.~J. Tulloch, and E.~G. Ritchie (2020).
\newblock A guide to ecosystem models and their environmental applications.
\newblock {\em Nature Ecology \& Evolution\/}~{\em 4\/}(11), 1459--1471.


\bibitem[\protect\citeauthoryear{Gellner, McCann, and Hastings}{Gellner et~al.}{2023}]{gellner_2023}
Gellner, G., K.~McCann, and A.~Hastings (2023).
\newblock Stable diverse food webs become more common when interactions are more biologically constrained.
\newblock {\em Proceedings of the National Academy of Sciences\/}~{\em 120\/}(31), e2212061120.


\bibitem[\protect\citeauthoryear{Gravel, Massol, and Leibold}{Gravel et~al.}{2016}]{gravel_2016_stabilitycomplexity}
Gravel, D., F.~Massol, and M.~A. Leibold (2016).
\newblock Stability and complexity in model meta-ecosystems.
\newblock {\em Nature Communications\/}~{\em 7\/}(1), 12457.


\bibitem[\protect\citeauthoryear{Grilli, Adorisio, Suweis, Barab{\'a}s, Banavar, Allesina, and Maritan}{Grilli et~al.}{2017}]{Grilli_2017_FS}
Grilli, J., M.~Adorisio, S.~Suweis, G.~Barab{\'a}s, J.~R. Banavar, S.~Allesina, and A.~Maritan (2017).
\newblock Feasibility and coexistence of large ecological communities.
\newblock {\em Nature Communications\/}~{\em 8\/}(1), 1--8.


\bibitem[\protect\citeauthoryear{Harfoot, Newbold, Tittensor, Emmott, Hutton, Lyutsarev, Smith, Scharlemann, and Purves}{Harfoot et~al.}{2014}]{harfoot_2014}
Harfoot, M.~B., T.~Newbold, D.~P. Tittensor, S.~Emmott, J.~Hutton, V.~Lyutsarev, M.~J. Smith, J.~P. Scharlemann, and D.~W. Purves (2014).
\newblock Emergent global patterns of ecosystem structure and function from a mechanistic general ecosystem model.
\newblock {\em PLoS biology\/}~{\em 12\/}(4), e1001841.


\bibitem[\protect\citeauthoryear{Hastings}{Hastings}{2001}]{Hastings_2001_transient}
Hastings, A. (2001).
\newblock Transient dynamics and persistence of ecological systems.
\newblock {\em Ecology Letters\/}~{\em 4\/}(3), 215--220.


\bibitem[\protect\citeauthoryear{Hastings, Abbott, Cuddington, Francis, Gellner, Lai, Morozov, Petrovskii, Scranton, and Zeeman}{Hastings et~al.}{2018}]{Hastings_2018_transient}
Hastings, A., K.~C. Abbott, K.~Cuddington, T.~Francis, G.~Gellner, Y.-C. Lai, A.~Morozov, S.~Petrovskii, K.~Scranton, and M.~L. Zeeman (2018).
\newblock Transient phenomena in ecology.
\newblock {\em Science\/}~{\em 361\/}(6406).


\bibitem[\protect\citeauthoryear{Holland}{Holland}{2023}]{holland_2023_thesis}
Holland, O. (2023).
\newblock {\em Assessing the risk of marine invasive species to nearshore marine ecosystems of {A}ustralia's {A}ntarctic research stations and subantarctic islands}.
\newblock Ph.\ D. thesis, Queensland University of Technology.

\bibitem[\protect\citeauthoryear{Hostetler, Kneip, Van~Vuren, and Oli}{Hostetler et~al.}{2012}]{hostetler_2012_stochastic}
Hostetler, J.~A., E.~Kneip, D.~H. Van~Vuren, and M.~K. Oli (2012).
\newblock Stochastic population dynamics of a montane ground-dwelling squirrel.
\newblock {\em PloS one\/}~{\em 7\/}(3), e34379.


\bibitem[\protect\citeauthoryear{Ives, Dennis, Cottingham, and Carpenter}{Ives et~al.}{2003}]{ives_2003}
Ives, A.~R., B.~Dennis, K.~L. Cottingham, and S.~R. Carpenter (2003).
\newblock Estimating community stability and ecological interactions from time-series data.
\newblock {\em Ecological monographs\/}~{\em 73\/}(2), 301--330.


\bibitem[\protect\citeauthoryear{Kristensen, Chisholm, and McDonald-Madden}{Kristensen et~al.}{2019}]{Kristensen_2019}
Kristensen, N.~P., R.~A. Chisholm, and E.~McDonald-Madden (2019).
\newblock Dealing with high uncertainty in qualitative network models using {B}oolean analysis.
\newblock {\em Methods in Ecology and Evolution\/}~{\em 10\/}(7), 1048--1061.


\bibitem[\protect\citeauthoryear{Kuhnert, Martin, and Griffiths}{Kuhnert et~al.}{2010}]{kuhnert_2010}
Kuhnert, P.~M., T.~G. Martin, and S.~P. Griffiths (2010).
\newblock A guide to eliciting and using expert knowledge in {B}ayesian ecological models.
\newblock {\em Ecology Letters\/}~{\em 13\/}(7), 900--914.


\bibitem[\protect\citeauthoryear{Landi, Minoarivelo, Br{\"a}nnstr{\"o}m, Hui, and Dieckmann}{Landi et~al.}{2018}]{landi_2018_review}
Landi, P., H.~O. Minoarivelo, {\AA}.~Br{\"a}nnstr{\"o}m, C.~Hui, and U.~Dieckmann (2018).
\newblock Complexity and stability of ecological networks: a review of the theory.
\newblock {\em Population Ecology\/}~{\em 60}, 319--345.


\bibitem[\protect\citeauthoryear{Ma'ayan}{Ma'ayan}{2017}]{ma_2017_systems}
Ma'ayan, A. (2017).
\newblock Complex systems biology.
\newblock {\em Journal of the Royal Society Interface\/}~{\em 14\/}(134), 20170391.


\bibitem[\protect\citeauthoryear{Martin, Burgman, Fidler, Kuhnert, Low-Choy, McBride, and Mengersen}{Martin et~al.}{2012}]{martin_2012}
Martin, T.~G., M.~A. Burgman, F.~Fidler, P.~M. Kuhnert, S.~Low-Choy, M.~McBride, and K.~Mengersen (2012).
\newblock Eliciting expert knowledge in conservation science.
\newblock {\em Conservation Biology\/}~{\em 26\/}(1), 29--38.


\bibitem[\protect\citeauthoryear{McDonald-Madden, Sabbadin, Game, Baxter, Chades, and Possingham}{McDonald-Madden et~al.}{2016}]{mcdonald_2016}
McDonald-Madden, E., R.~Sabbadin, E.~T. Game, P.~W. Baxter, I.~Chades, and H.~P. Possingham (2016).
\newblock Using food-web theory to conserve ecosystems.
\newblock {\em Nature Communications\/}~{\em 7\/}(1), 10245.


\bibitem[\protect\citeauthoryear{Melbourne-Thomas, Johnson, Fung, Seymour, Ch{\'e}rubin, Arias-Gonz{\'a}lez, and Fulton}{Melbourne-Thomas et~al.}{2011}]{melbourne_2011_spatial}
Melbourne-Thomas, J., C.~R. Johnson, T.~Fung, R.~M. Seymour, L.~M. Ch{\'e}rubin, J.~E. Arias-Gonz{\'a}lez, and E.~A. Fulton (2011).
\newblock Regional-scale scenario modeling for coral reefs: a decision support tool to inform management of a complex system.
\newblock {\em Ecological Applications\/}~{\em 21\/}(4), 1380--1398.


\bibitem[\protect\citeauthoryear{Monsalve-Bravo, Lawson, Drovandi, Burrage, Brown, Baker, Vollert, Mengersen, McDonald-Madden, and Adams}{Monsalve-Bravo et~al.}{2022}]{monsalve_2022_systems}
Monsalve-Bravo, G.~M., B.~A. Lawson, C.~Drovandi, K.~Burrage, K.~S. Brown, C.~M. Baker, S.~A. Vollert, K.~Mengersen, E.~McDonald-Madden, and M.~P. Adams (2022).
\newblock Analysis of sloppiness in model simulations: {U}nveiling parameter uncertainty when mathematical models are fitted to data.
\newblock {\em Science advances\/}~{\em 8\/}(38), eabm5952.


\bibitem[\protect\citeauthoryear{Moore, Wallington, Hobbs, Ehrlich, Holling, Levin, Lindenmayer, Pahl-Wostl, Possingham, Turner, et~al.}{Moore et~al.}{2009}]{moore_2009}
Moore, S.~A., T.~J. Wallington, R.~J. Hobbs, P.~R. Ehrlich, C.~Holling, S.~Levin, D.~Lindenmayer, C.~Pahl-Wostl, H.~Possingham, M.~G. Turner, et~al. (2009).
\newblock Diversity in current ecological thinking: implications for environmental management.
\newblock {\em Environmental management\/}~{\em 43}, 17--27.


\bibitem[\protect\citeauthoryear{Mori}{Mori}{2011}]{Mori_2011}
Mori, A.~S. (2011).
\newblock Ecosystem management based on natural disturbances: hierarchical context and non-equilibrium paradigm.
\newblock {\em Journal of Applied Ecology\/}~{\em 48\/}(2), 280--292.


\bibitem[\protect\citeauthoryear{Morozov, Abbott, Cuddington, Francis, Gellner, Hastings, Lai, Petrovskii, Scranton, and Zeeman}{Morozov et~al.}{2020}]{morozov_2020}
Morozov, A., K.~Abbott, K.~Cuddington, T.~Francis, G.~Gellner, A.~Hastings, Y.-C. Lai, S.~Petrovskii, K.~Scranton, and M.~L. Zeeman (2020).
\newblock Long transients in ecology: {T}heory and applications.
\newblock {\em Physics of Life Reviews\/}~{\em 32}, 1--40.


\bibitem[\protect\citeauthoryear{Mouquet, Lagadeuc, Devictor, Doyen, Duputi{\'e}, Eveillard, Faure, Garnier, Gimenez, Huneman, et~al.}{Mouquet et~al.}{2015}]{Mouquet_2015}
Mouquet, N., Y.~Lagadeuc, V.~Devictor, L.~Doyen, A.~Duputi{\'e}, D.~Eveillard, D.~Faure, E.~Garnier, O.~Gimenez, P.~Huneman, et~al. (2015).
\newblock Predictive ecology in a changing world.
\newblock {\em Journal of Applied Ecology\/}~{\em 52\/}(5), 1293--1310.


\bibitem[\protect\citeauthoryear{Newsome, Ballard, Crowther, Dellinger, Fleming, Glen, Greenville, Johnson, Letnic, Moseby, et~al.}{Newsome et~al.}{2015}]{newsome_2015_dingo}
Newsome, T.~M., G.-A. Ballard, M.~S. Crowther, J.~A. Dellinger, P.~J. Fleming, A.~S. Glen, A.~C. Greenville, C.~N. Johnson, M.~Letnic, K.~E. Moseby, et~al. (2015).
\newblock Resolving the value of the dingo in ecological restoration.
\newblock {\em Restoration Ecology\/}~{\em 23\/}(3), 201--208.


\bibitem[\protect\citeauthoryear{Novak, Wootton, Doak, Emmerson, Estes, and Tinker}{Novak et~al.}{2011}]{Novak_2011}
Novak, M., J.~T. Wootton, D.~F. Doak, M.~Emmerson, J.~A. Estes, and M.~T. Tinker (2011).
\newblock Predicting community responses to perturbations in the face of imperfect knowledge and network complexity.
\newblock {\em Ecology\/}~{\em 92\/}(4), 836--846.


\bibitem[\protect\citeauthoryear{Oro and Mart{\'\i}nez-Abra{\'\i}n}{Oro and Mart{\'\i}nez-Abra{\'\i}n}{2023}]{oro_2023}
Oro, D. and A.~Mart{\'\i}nez-Abra{\'\i}n (2023).
\newblock Ecological non-equilibrium and biological conservation.
\newblock {\em Biological Conservation\/}~{\em 286}, 110258.


\bibitem[\protect\citeauthoryear{Pascal}{Pascal}{2024}]{Pascal_2024_code}
Pascal, L.~V. (2024).
\newblock {EEMtoolbox}.
\newblock \url{https://github.com/luzvpascal/EEMtoolbox}.

\bibitem[\protect\citeauthoryear{Pesendorfer, Baker, Stringer, McDonald-Madden, Bode, McEachern, Morrison, and Sillett}{Pesendorfer et~al.}{2018}]{Pesendorfer_2018_egEEM}
Pesendorfer, M.~B., C.~M. Baker, M.~Stringer, E.~McDonald-Madden, M.~Bode, A.~K. McEachern, S.~A. Morrison, and T.~S. Sillett (2018).
\newblock Oak habitat recovery on {C}alifornia's largest islands: scenarios for the role of corvid seed dispersal.
\newblock {\em Journal of Applied Ecology\/}~{\em 55\/}(3), 1185--1194.


\bibitem[\protect\citeauthoryear{Peterson and Bode}{Peterson and Bode}{2021}]{Peterson_2021}
Peterson, K. and M.~Bode (2021).
\newblock Using ensemble modeling to predict the impacts of assisted migration on recipient ecosystems.
\newblock {\em Conservation Biology\/}~{\em 35\/}(2), 678--687.


\bibitem[\protect\citeauthoryear{Peterson, Barnes, Jeynes-Smith, Cowen, Gibson, Sims, Baker, and Bode}{Peterson et~al.}{2021}]{Peterson_2021_DirkHartog}
Peterson, K.~A., M.~D. Barnes, C.~Jeynes-Smith, S.~Cowen, L.~Gibson, C.~Sims, C.~M. Baker, and M.~Bode (2021).
\newblock Reconstructing lost ecosystems: A risk analysis framework for planning multispecies reintroductions under severe uncertainty.
\newblock {\em Journal of Applied Ecology\/}~{\em 58\/}(10), 2171--2184.


\bibitem[\protect\citeauthoryear{Plein, O'Brien, Holden, Adams, Baker, Bean, Sisson, Bode, Mengersen, and McDonald-Madden}{Plein et~al.}{2022}]{plein_2022}
Plein, M., K.~R. O'Brien, M.~H. Holden, M.~P. Adams, C.~M. Baker, N.~G. Bean, S.~A. Sisson, M.~Bode, K.~L. Mengersen, and E.~McDonald-Madden (2022).
\newblock Modeling total predation to avoid perverse outcomes from cat control in a data-poor island ecosystem.
\newblock {\em Conservation Biology\/}~{\em 36\/}(5), e13916.


\bibitem[\protect\citeauthoryear{Possingham, Andelman, Noon, Trombulak, and Pulliam}{Possingham et~al.}{2001}]{Possingham_2001}
Possingham, H.~P., S.~Andelman, B.~Noon, S.~Trombulak, and H.~Pulliam (2001).
\newblock Making smart conservation decisions.
\newblock {\em Conservation Biology: research priorities for the next decade\/}~{\em 23}, 225--244.


\bibitem[\protect\citeauthoryear{Rayner, Hauber, Imber, Stamp, and Clout}{Rayner et~al.}{2007}]{rayner_2007_spatial}
Rayner, M.~J., M.~E. Hauber, M.~J. Imber, R.~K. Stamp, and M.~N. Clout (2007).
\newblock Spatial heterogeneity of mesopredator release within an oceanic island system.
\newblock {\em Proceedings of the National Academy of Sciences\/}~{\em 104\/}(52), 20862--20865.


\bibitem[\protect\citeauthoryear{Reimer, Arroyo-Esquivel, Jiang, Scharf, Wolkovich, Zhu, and Boettiger}{Reimer et~al.}{2021}]{reimer_2021}
Reimer, J., J.~Arroyo-Esquivel, J.~Jiang, H.~Scharf, E.~Wolkovich, K.~Zhu, and C.~Boettiger (2021).
\newblock Noise can create or erase long transient dynamics.
\newblock {\em Theoretical Ecology\/}~{\em 14\/}(4), 685--695.


\bibitem[\protect\citeauthoryear{Rendall, Sutherland, Baker, Raymond, Cooke, and White}{Rendall et~al.}{2021}]{Rendall_2021_EEMeg}
Rendall, A.~R., D.~R. Sutherland, C.~M. Baker, B.~Raymond, R.~Cooke, and J.~G. White (2021).
\newblock Managing ecosystems in a sea of uncertainty: invasive species management and assisted colonizations.
\newblock {\em Ecological Applications\/}~{\em 31\/}(4), e02306.


\bibitem[\protect\citeauthoryear{Rohr, Saavedra, and Bascompte}{Rohr et~al.}{2014}]{Rohr_2014}
Rohr, R.~P., S.~Saavedra, and J.~Bascompte (2014).
\newblock On the structural stability of mutualistic systems.
\newblock {\em Science\/}~{\em 345\/}(6195).


\bibitem[\protect\citeauthoryear{S{\ae}ther, Engen, Filli, Aanes, Schr{\"o}der, and Andersen}{S{\ae}ther et~al.}{2002}]{saether_2002_stochastic}
S{\ae}ther, B.-E., S.~Engen, F.~Filli, R.~Aanes, W.~Schr{\"o}der, and R.~Andersen (2002).
\newblock Stochastic population dynamics of an introduced swiss population of the ibex.
\newblock {\em Ecology\/}~{\em 83\/}(12), 3457--3465.


\bibitem[\protect\citeauthoryear{Schlick}{Schlick}{2010}]{schlick_2010_molecular}
Schlick, T. (2010).
\newblock {\em Molecular modeling and simulation: an interdisciplinary guide}, Volume~2.
\newblock Springer.


\bibitem[\protect\citeauthoryear{Song, Rohr, and Saavedra}{Song et~al.}{2018}]{song_2018_feasibility}
Song, C., R.~P. Rohr, and S.~Saavedra (2018).
\newblock A guideline to study the feasibility domain of multi-trophic and changing ecological communities.
\newblock {\em Journal of Theoretical Biology\/}~{\em 450}, 30--36.


\bibitem[\protect\citeauthoryear{Sunn{\aa}ker, Busetto, Numminen, Corander, Foll, and Dessimoz}{Sunn{\aa}ker et~al.}{2013}]{sunnaaker_2013_ABC}
Sunn{\aa}ker, M., A.~G. Busetto, E.~Numminen, J.~Corander, M.~Foll, and C.~Dessimoz (2013).
\newblock Approximate {B}ayesian computation.
\newblock {\em {PLoS Computational Biology}\/}~{\em 9\/}(1), e1002803.


\bibitem[\protect\citeauthoryear{Sutherland}{Sutherland}{2006}]{sutherland_2006}
Sutherland, W.~J. (2006).
\newblock Predicting the ecological consequences of environmental change: a review of the methods.
\newblock {\em Journal of Applied Ecology\/}, 599--616.


\bibitem[\protect\citeauthoryear{Tulloch, Hagger, and Greenville}{Tulloch et~al.}{2020}]{tulloch_2020}
Tulloch, A.~I., V.~Hagger, and A.~C. Greenville (2020).
\newblock Ecological forecasts to inform near-term management of threats to biodiversity.
\newblock {\em Global Change Biology\/}~{\em 26\/}(10), 5816--5828.


\bibitem[\protect\citeauthoryear{Vollert, Drovandi, and Adams}{Vollert et~al.}{2024a}]{Vollert_2024_code}
Vollert, S.~A., C.~Drovandi, and M.~P. Adams (2024a).
\newblock Code repository for {U}nlocking ensemble ecosystem modelling for large and complex networks.
\newblock \url{https://figshare.com/articles/software/Code_repository_for_Unlocking_ensemble_ecosystem_modelling_for_large_and_complex_networks_/23707119}.

\bibitem[\protect\citeauthoryear{Vollert, Drovandi, and Adams}{Vollert et~al.}{2024b}]{vollert_2023_SMCEEM}
Vollert, S.~A., C.~Drovandi, and M.~P. Adams (2024b).
\newblock Unlocking ensemble ecosystem modelling for large and complex networks.
\newblock {\em PLoS Computational Biology\/}~{\em 20\/}(3), e1011976.


\bibitem[\protect\citeauthoryear{Wallington, Hobbs, and Moore}{Wallington et~al.}{2005}]{wallington_2005}
Wallington, T.~J., R.~J. Hobbs, and S.~A. Moore (2005).
\newblock Implications of current ecological thinking for biodiversity conservation: a review of the salient issues.
\newblock {\em Ecology and Society\/}~{\em 10\/}(1).


\bibitem[\protect\citeauthoryear{Waltner-Toews, Kay, Neudoerffer, and Gitau}{Waltner-Toews et~al.}{2003}]{waltner_2003}
Waltner-Toews, D., J.~J. Kay, C.~Neudoerffer, and T.~Gitau (2003).
\newblock Perspective changes everything: managing ecosystems from the inside out.
\newblock {\em Frontiers in Ecology and the Environment\/}~{\em 1\/}(1), 23--30.


\end{thebibliography}
